\documentclass[preprint]{aastex}
\usepackage{graphicx}
\usepackage{times}
\newcommand{\OTwo}{[\ion{O}{2}]$\lambda 3727$}
\newcommand{\OThree}{[\ion{O}{3}]$\lambda 5007$}

\begin{document}

\bibliographystyle{apj}

\slugcomment{To appear in The Astronomical Journal}

\title{The Gemini Deep Deep Survey: I. Introduction to the Survey, 
Catalogs and Composite Spectra}

\author{\medskip Roberto G. Abraham}
\affil{
   Department of Astronomy \& Astrophysics,
   University of Toronto, 60 St. George St,
   Toronto, ON, M5S~3H8, Canada.
}
\email{abraham@astro.utoronto.ca}

\author{Karl Glazebrook}
\affil{
  Department of Physics \& Astronomy,
  Johns Hopkins University,
  3400 North Charles Street,
  Baltimore, MD 21218-2686.
}
\email{kgb@pha.jhu.edu}

\author{Patrick J. McCarthy}
\affil{
 Observatories of the Carnegie Institution of Washington,
 813 Santa Barbara Street,
 Pasadena, CA 91101.
}
\email{pmc2@ociw.edu}

\author{David Crampton, Richard Murowinski} 
\affil{Herzberg Institute of Astrophysics,
5071 West Saanich Road, Victoria,
British Columbia, V9E~2E7, Canada.} 
\email{david.crampton@nrc.ca, richard.murowinski@nrc.ca} 

\author{Inger J{\o}rgensen, Kathy Roth}
\affil{
  Gemini Observatory, 
  Hilo, HI 96720
}
\email{ijorgensen@gemini.edu, kroth@gemini.edu}

\author{Isobel M. Hook}
\affil{
 Department of Astrophysics,
 Oxford University,
 Keble Road,
 Oxford OX1 3RH,
 England.
}
\email{imh@astro.ox.ac.uk}

\author{Sandra Savaglio\altaffilmark{1}}
\affil{
  Department of Physics \& Astronomy,
  Johns Hopkins University,
  3400 North Charles Street,
  Baltimore, MD 21218-2686.
}
\altaffiltext{1}{On leave of absence from INAF, Osservatorio Astronomico di Roma, Italy.}
\email{savaglio@pha.jhu.edu}  

\author{Hsiao-Wen Chen\altaffilmark{2}}
\affil{
  Center for Space Research, 
  Massachusetts Institute of Technology, 
  Cambridge, MA 02139-4307.
}
\altaffiltext{2}{Hubble Fellow}
\email{hchen@space.mit.edu}

\author{Ronald O. Marzke}
\affil{
  Dept. of Physics and Astronomy,
  San Francisco State University,
  1600 Holloway Avenue, 
  San Francisco, CA 94132 .
}
\email{marzke@stars.sfsu.edu}

\author{R. G. Carlberg}
\affil{
   Department of Astronomy \& Astrophysics,
   University of Toronto, 60 St. George St,
   Toronto ON, M5S~3H8, Canada.
}
\email{carlberg@astro.utoronto.ca}

\begin{abstract} 
  The Gemini Deep Deep Survey (GDDS) is an ultra-deep ($K<20.6$ mag,
  $I<24.5$ mag) redshift survey targeting galaxies in the ``redshift
  desert'' between $1<z<2$. The primary goal of the survey is to
  constrain the space density at high redshift of evolved high-mass
  galaxies. We obtained 309 spectra in four widely-separated 30
  arcmin$^2$ fields using the Gemini North telescope and the Gemini
  Multi-Object Spectrograph (GMOS). The spectra define a one-in-two
  sparse sample of the reddest and most luminous galaxies near the
  $I-K$\, vs. $I$\, color-magnitude track mapped out by passively
  evolving galaxies in the redshift interval $0.8<z<1.8$.  This sample
  is augmented by a one-in-seven sparse sample of the remaining
  high-redshift galaxy population.  The GMOS spectrograph was
  operating in a Nod \& Shuffle mode which enabled us to remove sky
  contamination with high precision, even for typical exposures times
  of 20--30 hours per field.  The resulting spectra are the deepest
  ever obtained.  In this paper we present our sample of 309 spectra,
  along with redshifts, identifications of spectral features, and
  photometry. This makes the GDDS the largest and most complete
  infrared-selected survey probing the redshift desert.  The 7-band
  ($VRIzJHK_s$) photometry is taken from the Las Campanas Infrared
  Survey.  The infrared selection means that the GDDS is observing not
  only star-forming galaxies, as in most high-redshift galaxy surveys,
  but also quiescent evolved galaxies.  In our sample, we have obtained 
  225 secure redshifts, 167 of which are in the redshift interval $0.8 < z < 2$. 
  About 25\% of these show clear spectral signatures of evolved (pure old, 
  or old + intermediate-age) stellar populations, while 35\% of show 
  features consistent with either a pure intermediate-age or a 
  young + intermediate-age stellar population. About 29\% of the 
  galaxies in the GDDS at $0.8 < z < 2$ are young starbursts with 
  strong interstellar lines. A few galaxies show very strong post-starburst 
  signatures. Another 55 objects have less secure redshifts, 31 of
  which lie in the redshift interval $0.8 < z < 2$.  The median
  redshift of the whole GDDS sample is $z=1.1$.  Spectroscopic
  completeness varies from a low of $\sim70\%$ for red galaxies to
  $>90\%$ for blue galaxies. In this paper we also present, together
  with the data and catalogs, a summary of the criteria for selecting
  the GDDS fields, the rationale behind our mask designs, an analysis
  of the completeness of the survey, and a description of the data
  reduction procedures used. All data from the GDDS are publicly
  available.
\end{abstract}


\section{INTRODUCTION}
\label{sec:introduction}


The Gemini Deep Deep Survey (GDDS) is an infrared-selected ultra-deep
spectroscopic survey probing the redshift range $0.8<z<1.8$. It is
designed to target galaxies of all colors at high redshift with an
emphasis on the reddest population.  The survey is designed with the
following scientific goals in mind: (1) Measurement of the space
density and luminosity function of massive early-type galaxies at high
redshift. (2) Construction of the volume-averaged stellar mass
function in at least three mass bins and two redshift bins over the
target redshift range. (3) Measurement of the luminosity-weighted ages
and recent star-formation histories of $\sim 50$ evolved galaxies at
$z>1$ \citep[cf.][]{dun96}. The over-arching goal of the survey is to
use these sets of observations to test hierarchical models for the
formation of early-type galaxies.  Many studies (see \citet{ell01} for
a recent review) have probed the evolving space density of early-type
systems and it is now clear that the number density of early-types
does not evolve rapidly out to $z=1$, as once predicted by
matter-dominated models (see \citealt {ell01} for a review).  However,
$\Lambda$-dominated cosmologies push back the formation epoch of most
early-type systems out to at least $z=1$ even in a hierarchical
picture.  Alternative theories for the origin of early-type galaxies
(e.g. high-z monolithic collapse {\em vs.} hierarchical formation from
mergers) now start to become readily distinguishable at exactly the
redshift ($z=1$) where spectroscopy from the ground becomes
problematic \citep{kau98}.

Our focus on the redshift range $0.8<z<1.8$ is motivated by two additional
considerations. Firstly, the star-formation histories of individual galaxies in
this redshift range have been very poorly explored.  We do not even know whether
most red objects in this range are old and quiescent, or very young and active
and heavily reddened by dust. Distinguishing between these two possibilities
requires high signal-to-noise in the continuum so that the characteristic
photospheric features of evolved stars become evident, but most work in this
redshift range has focused on emission lines. Secondly, and irrespective of
model predictions, this redshift range appears to correspond to the peak epoch
of galaxy assembly inferred by integrating under the `Madau/Lilly plot',
an observationally-defined diagram quantifying the volume-averaged
star-formation history of the Universe as a function of redshift \citep{mad96,lil96,ste99}.
The high-redshift tail of this plot is subject to large and uncertain dust and
surface-brightness corrections, and remains rather poorly determined, and recent observations have pushed back the peak of star-formation, showing a broad maximum in redshift (for
a summary of the observational situation, see Figure 2 in 
Nagamine et al. 2003). However,
the integral under the Madau/Lilly plot is simply the total mass assembled in
stars per unit volume, so by measuring this quantity directly in the GDDS we can
undertake a basic consistency check of the overall picture inferred from global
star-formation history and luminosity density diagrams.

Spectroscopy of galaxies in the redshift range we seek to probe
suffers from technical challenges brought on by the lack of strong
spectral features at visible wavelengths. The redshift range $1<z<2$
has come to be known as the ``redshift desert'', in reference to the
paucity of optical redshifts known over this interval. Fortunately, it
is now becoming clear that redshifts and diagnostic spectra {\em can}
be obtained using optical spectrographs in this redshift range, by
focusing on rest-frame UV metallic absorption features. Good progress
is now being made in obtaining redshifts for UV-selected samples in
the redshift desert using the blue-sensitive LRIS-B spectrograph on
the Keck telescope \citep{ste03,erb03}. However, UV-selected surveys
are biased in favor of high star-formation rate galaxies, and the
passive red galaxies with high mass-to-light ratios that are missed by
UV-selection could well dominate the high-$z$ galaxy mass budget,
motivating deep $K$-selected surveys such as the VLT K20 survey
\citep{cim03}, and ultra-deep small area surveys such as FIRES
\citep{fra03}. We refer the reader to \citet{cim04} for an excellent
summary of recent results obtained from infrared-selected surveys
probing high-redshift galaxy evolution.  The GDDS is designed to build
upon these results.

Because their rest-UV continuum is so weak, determining the redshifts
of passive red galaxies at $z\sim 1.5$ with 8m-class telescopes
presently requires extreme measures. Ultra-deep ($>10$ hour)
integration times and Poisson-limited spectroscopy are required in
order to probe samples of red galaxies with zero residual
star-formation and no emission lines.  This poses a severe problem,
because MOS spectroscopy with 8m-class telescopes is generally not
Poisson-limited unless exposure times are short (less than a few
hours). The main contributors to the noise budget are imperfect sky
subtraction and fringe removal. At optical wavelengths, both of these
problems are most severe redward of 7000\AA, where most of the light
from evolved high redshift stellar populations is expected to peak. To
mitigate against these effects, the Gemini Deep Deep Survey team has
implemented a Nod \& Shuffle sky-subtraction mode \citep{gla01,cui94,
  bla95} on the Gemini Multi-Object Spectrograph \citep{mur03,hoo03}.
This technique is somewhat similar to beam-switching in the infrared,
and allows sky subtraction and fringe removal to be undertaken with an
order of magnitude greater precision than is possible with
conventional spectroscopy.

In order to undertake an unbiased inventory of the high-redshift
galaxy population, the GDDS is $K$-band selected to a sufficient depth
($K = 20.6$ mag) to reach $L^\star$ throughout the $1 < z < 2$ regime.
IR-selected samples that do not reach to this $K$-band limit are
limited primarily to the $z < 1$ epoch, while samples that go
substantially deeper outrun the capability of ground-based
spectroscopic follow-up for the reddest objects, and once again become
biased. The standard definition for an `Extremely Red Galaxy', or ERG,
is $I-K\ga 4$, a threshold which roughly corresponds to the expected
color of an evolved dust-free early-type galaxy seen at $z\sim 1$. As
will be shown below, the effective limit for obtaining absorption-line
redshifts for red galaxies with weak UV continua is about $I=25$ mag
with an 8m telescope. (Our GDDS spectroscopy --- the deepest ever
undertaken --- has a magnitude limit of $I=24.5$ mag). Therefore at
present it is only just possible to obtain a nearly complete census of
redshifts for all evolved red objects in a $K\sim21$ imaging survey.
Our strategy with the GDDS is to go deep enough to allow redshifts to
be obtained for $L_\star$ galaxies irrespective of star-formation
history at $z\sim 1.5$, while simultaneously covering enough area to
minimize the effects of cosmic variance.  Our sampling strategy (based
on photometric redshift pre-selection to eliminate low-$z$
contamination) is different from that adopted by most other redshift
surveys. In terms of existing surveys, the K20 survey \citep{cim03} is
probably the closest benchmark comparison to the GDDS, although the
experimental designs are very different, making the K20 and GDDS
surveys quite complementary.  The K20 survey has about twice as many
redshifts as the GDDS, but because K20 survey does not preferentially
select against low-redshift objects, most of these are at $z<1$. The
GDDS has between two and three times as many redshifts as K20 in the
interval $1.2<z<2$ (the precise number depending on the minimum
acceptable redshift confidence class), and has a higher median
redshift ($z\sim 1.1$ vs. $z\sim0.7$).  The GDDS also goes about 0.6
mag deeper in $K$ and has over twice the area (121 square arcmin in
four widely-separated site-lines in the GDDS vs. 52 square arcmin in
two widely-separated site-lines in K20).

A plan for this paper follows. In \S2 we describe our experimental
design, with a particular focus on how our targets were selected from
the {\em Las Campanas Infrared Survey}.  In \S3 we outline our
observing procedure, but defer the details of the Nod \& Shuffle mode
that are not specific to the GDDS to an Appendix. (Nod \& Shuffle on
the Gemini Multi-Object Spectrograph was implemented for use on the
GDDS but is now a common-user mode. Since many observers may wish to
use the mode themselves in contexts unrelated to faint galaxy
observations, the Appendices to this paper will act as a stand-alone
reference to using the mode on Gemini).  In Section 4 we summarize the
data obtained from the GDDS, both graphically and as a series of
tables. Composite spectra obtained by co-adding similar spectra are
presented in Section 5. We used these composite as templates to obtain
redshifts in the GDDS, but others may wish to apply them to their own
work for other purposes\footnote{It should be born in mind that these
  composites are constructed from galaxies covering a wide range of
  redshift and time. Most analyses on composites would require a more
  restricted range, e.g.  paper II (Savaglio et al. 2004).}.  Some
implications from the data obtained are discussed and our conclusions
given in Section 6. The major results from the GDDS will be presented
in three companion papers\footnote{Savaglio et al. 2004 [paper II]
  presents measurements of column densities and metallicities of
  star-forming galaxies in our Sample. Glazebrook et al. 2004 [paper
  III] presents the mass function from the GDDS. McCarthy et al. 2004
  [paper IV] will present an analysis of the stellar populations in
  the reddest galaxies in our sample.}.

Appendix A describes the operation of the Nod \& Shuffle mode on the
Gemini Multi-Object Spectrograph \citep{hoo03} (GMOS) in the context
of the GDDS. A more general description of the implementation of the
mode will be given in Murowinski et al. (in preparation). Appendix B
describes how the two-dimensional data from the GDDS were reduced.
Appendix C describes how the final one-dimensional spectra were
extracted from the two-dimensional data.

The catalogs presented in this paper, as well as reduced spectra for
all galaxies in the GDDS, are available in electronic form as a
digital supplement to this article. All software described in this
paper, as well as auxiliary data, are publicly available (in both raw
and fully reduced form) from the central GDDS web site located at
http://www.ociw.edu/lcirs/gdds.html.

Throughout this paper we adopt a cosmology with $H_0$=70 km/s/Mpc,
$\Omega_M=0.3$, and $\Omega_\Lambda=0.7$.

\section{EXPERIMENTAL DESIGN}
\label{sec:design}

\begin{figure*}[htbp]
\begin{center}
\includegraphics[width=6in]{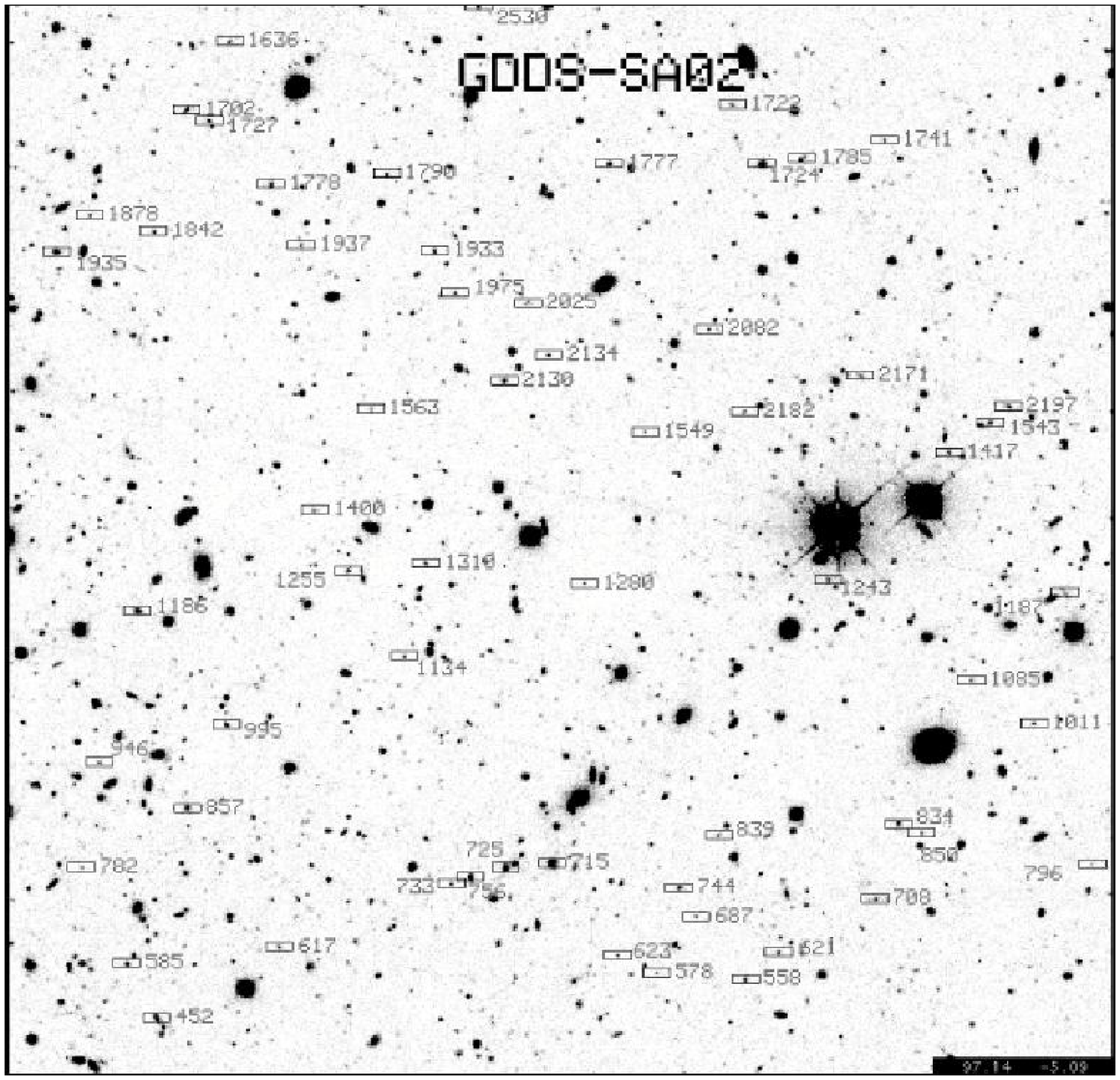} 
\caption{\label{fig:chart1} 
  Labeled finding chart for the GDDS-SA02 field. The field size is
  5.5\arcmin x 5.5\arcmin. The central coordinates for this field are
  given in Table~\ref{tab:overview}. Numeric object labels correspond
  to the ID's in Tables 4 and 5. The background image is a 180 min
  $I$-band image obtained with the KPNO 4m MOSIAC imager as part of
  the LCIR survey.  }
\end{center}
\end{figure*}

\begin{figure*}[htbp]
\begin{center}
\includegraphics[width=6in]{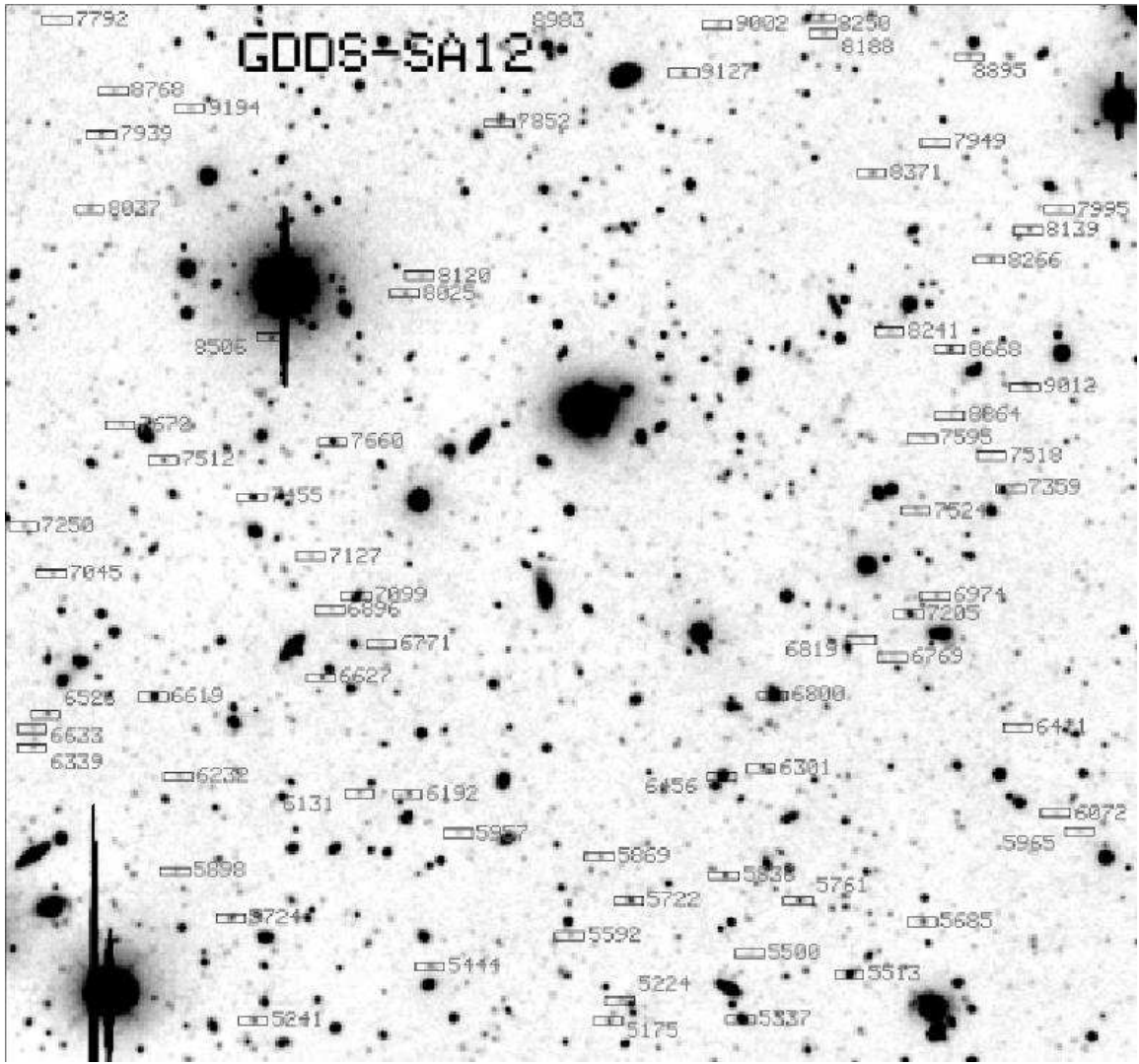} 
\caption{\label{fig:chart2}  Labeled finding chart for the GDDS-SA12
  field. The field size is 5.5\arcmin x 5.5\arcmin. Central
  coordinates for this field are given in Table~\ref{tab:overview}.
  Numeric object labels correspond to the ID's in Tables 4 and 5. The
  background image is a 180 minute $I$-band exposure taken with the
  BTC on the CTIO 4m telescope as part of the LCIR imaging survey.}
\end{center}
\end{figure*}

\begin{figure*}[htbp]
\begin{center}
\includegraphics[width=6in]{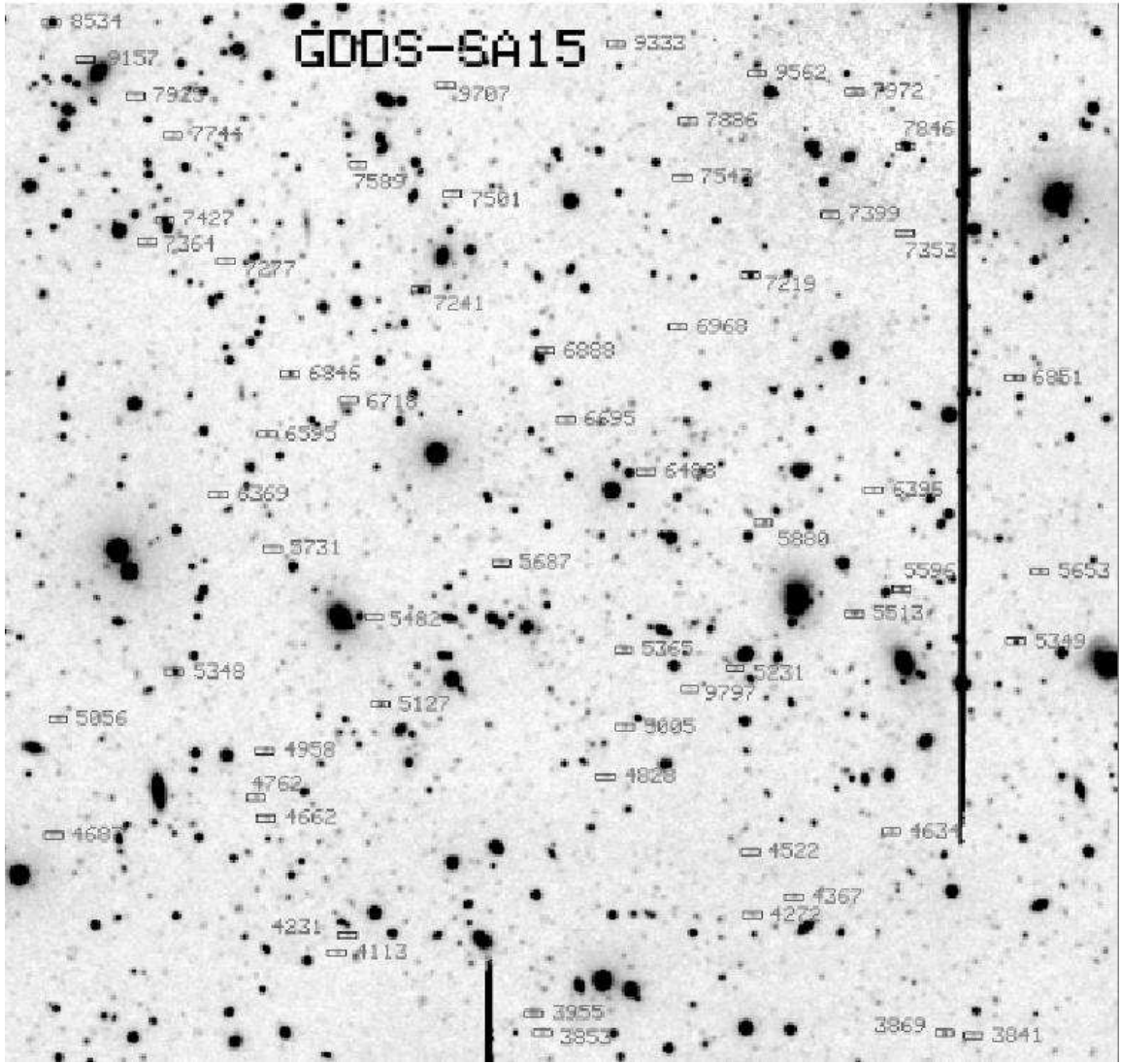} 
\caption{\label{fig:chart3}  Labeled finding chart for the GDDS-SA15
  field. The field size is 5.5\arcmin x 5.5\arcmin. Central
  coordinates for this field are given in Table~\ref{tab:overview}.
  Numeric object labels correspond to the ID's in Tables 4 and 5. The
  background is a 180 minute $I$-band image obtained with the BTC on
  the CTIO 4m as part of the LCIR survey.}
\end{center}
\end{figure*}

\begin{figure*}[htbp]
\begin{center}
\includegraphics[width=6in]{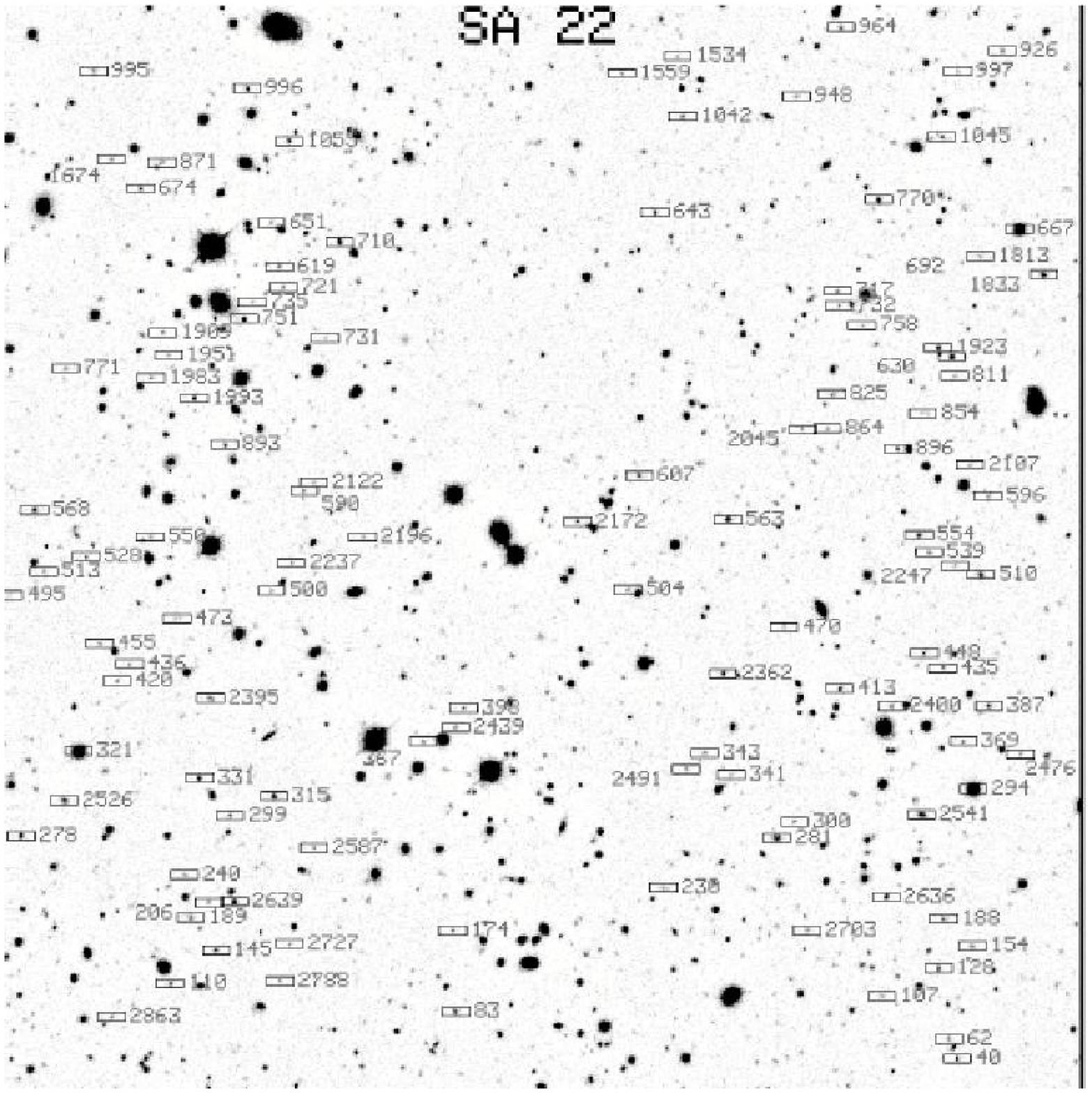} 
\caption{\label{fig:chart4}  Labeled finding chart for the GDDS-SA22
  field. The field size is 5.5\arcmin x 5.5\arcmin. Central
  coordinates for this field are given in Table~\ref{tab:overview}.
  Numeric object labels correspond to the ID's in Tables 4 and 5. The
  background image is a 120 minute $I$-band exposure obtained with the
  12k mosaic camera on the CFHT kindly provided to us by P. Stetson.}
\end{center}
\end{figure*}

All galaxies observed in the GDDS were taken from seven-filter
($BVRIz^\prime JK$) photometric catalogs constructed as part of the
one square-degree Las Campanas Infrared survey (LCIR survey;
\citealp{mcc01,che02,fir01}). The GDDS can be thought of as a
sparse-sampled spectroscopically defined subset of the LCIR survey.
The GDDS is comprised of four Gemini Multi-Object Spectrograph (GMOS)
integrations, each with exposure times between 21 and 38 hours. Each
GDDS field lies within a separate LCIR survey equatorial field.  (The
LCIR survey fields chosen were SSA22, NOAODW-Cetus, NTT-Deep, and LCIR
1511; the reader is referred to \citealt{che02} for information on the
LCIR survey data available in these fields). Since the publication of
\citealt{che02} additional imaging data has been acquired for these
fields. Deep $K_s$, and in some cases, J imaging supplements the
VVRIz'H data discussed in Chen et al. The 5$\sigma$ completeness limit
of the LCIR survey fields is K$_s = 20.6$ mag (on the Vega scale).
The 5.5\arcmin x 5.5\arcmin\ GMOS field of view is small relative to
even a single 13\arcmin x 13\arcmin\ `tile' of the LCIR survey (four
such tiles constitute a single LCIR survey field). We therefore had
considerable freedom to position the GMOS pointing within each LCIR
survey field in areas which avoided very bright foreground objects and
which had suitable guide stars proximate to the fields. We were also
careful to pick regions in each field where the number of red galaxies
was near the global average ({\em i.e.} we tried to avoid obvious
over-densities and obvious voids; our success in achieving this will
be quantified below). The four GMOS fields in our survey will be
referred to as GDDS-SA22, GDDS-SA02, GDDS-SA12 and GDDS-SA15 for the
remainder of this paper and in subsequent papers in this series. Taken
together, these survey fields cover a total area of 121 square arcmin.
The locations of each field and the total exposure time per
spectroscopic mask are given in Table~\ref{tab:overview}. Finding
charts for individual galaxies within the GDDS fields are shown in
Figures~\ref{fig:chart1}--\ref{fig:chart4}.

\begin{deluxetable}{cccccc}
\tablecaption{Overview of Observations\label{tab:overview}}
\tablecolumns{6}
\tablewidth{0pc}
\tabletypesize{\small}
\tablehead{
  \colhead{Field} &
  \colhead{RA (J2000)} &
  \colhead{Dec (J2000)} &
  \colhead{Target slits} &
  \colhead{Mask Definition File\tablenotemark{a}} &
  \colhead{Exposure Time(s)}
}
\startdata
GDDS-SA02 & 02:09:41.30 &  -04:37:54.0 & 59 & GN2002B-Q-1-1.fits   & 75600 \\
GDDS-SA12 & 12:05:22.17 & -07:22:27.9  &  61 & GN2003A-Q-1-1.fits & 18000\tablenotemark{b} \\
          "             & 12:05:22.17 & -07:22:27.9  &  74 & GN2003A-Q-1-3.fits & 57600\tablenotemark{b} \\
GDDS-SA15 & 15:23:47.83  & -00:05:21.1  & 59 & GN2003A-Q-1-5.fits  & 70200 \\
GDDS-SA22 & 22:17:41.0 & +00:15:20.0  & 83 &  GN2002BSV-78-14.fits & 48600\tablenotemark{c} \\
          "             & 22:17:41.0 & +00:15:20.0  & 62 &  GN2002BSV-78-15.fits & \ 90000\tablenotemark{c} \enddata
\tablenotetext{a}{\ FITS-format mask definition file stored the Gemini public data archive.}
\tablenotetext{b}{\ 48 slits were in common between masks GN2003A-Q-1-1.fits and
GN2003A-Q-1-3.fits. These objects have a total exposure time of 75600s.}
\tablenotetext{c}{\ 27 slits were in common between masks GN2002BSV-78-14 and
 GN2002BSV-78-15. These objects have a total exposure time of 138600s.}
\end{deluxetable}

\begin{figure*}[htbp]
\begin{center}
\includegraphics[width=7in]{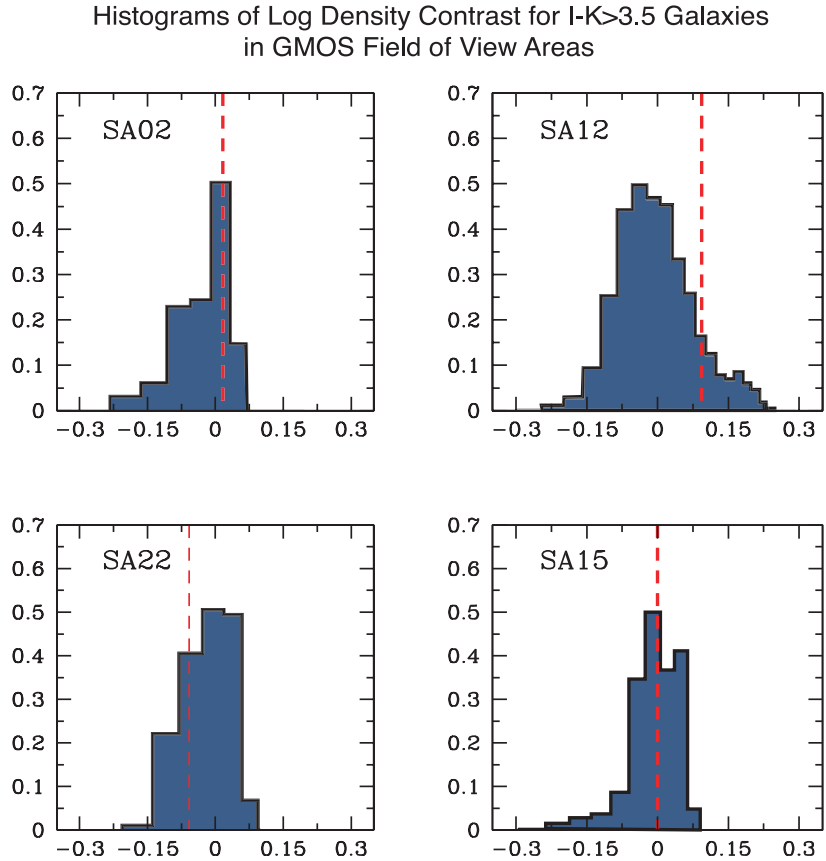} 
\caption{\label{fig:selection}
The density contrast of $(I-K)>4$ galaxies in each 5.5 arcmin x 5.5
arcmin GDDS field (dashed vertical line) compared with the histogram
obtained by measuring this quantity in a series of random boxes overlaid
on the \hbox{26\arcmin\ x 26\arcmin}~ LCIRS field from which the GMOS
field was selected.}
\end{center}
\end{figure*}

At $z=1.2$ (the median redshift of our survey), the 5.5 arcmin angle
subtended by each GMOS field of view corresponds to a physical size of
2.74 Mpc. The total comoving volume in the four GDDS `pencil beams'
over the redshift interval $0.8<z<1.8$ (the range over which $L^\star$
galaxies would be detected in the GDDS) is 320,000 ${\rm Mpc}^3$.
Over this volume the effects of cosmic variance on random pointings is
quite significant, especially for the highly clustered red objects in
our survey whose correlation length is large ($\sim 10h^{-1}$ Mpc;
\citealt{mcc01,dad00}). Fortunately, completely random pointings are
unnecessary, because the global statistical properties of the the LCIR
survey (the GDDS' parent population) are well defined. As mentioned
earlier, we took advantage of this extra information when selecting
our GDDS fields by ensuring that the areal density of red ($I-K>4$)
galaxies was close to the ensemble average of such galaxies in the
LCIR survey.  Figure~\ref{fig:selection} compares the density contrast
of red galaxies in each of our fields to the histogram obtained by
measuring this same quantity in a series of random 5.5\arcmin\ x
5.5\arcmin\ boxes overlaid on the \hbox{26\arcmin\ x 26\arcmin}~ LCIRS
field from which the corresponding GDDS field was selected.  In
GDDS-SA15 and GDDS-SA02 the number of red galaxies is essentially
identical to the median number expected from the parent population.
The number of red galaxies in GDDS-SA22 is somewhat lower than the
global median, while the number in GDDS-SA12 is somewhat higher than
the global median. Averaged over the four fields, the areal density of
red systems in the GDDS is close to the typical value expected.

The areal density of galaxies in the LCIR Survey with photometric
redshifts in the range $1<z<2$ and $I<24.5$ mag is about 8
arcmin$^{-2}$, corresponding to about 250 galaxies in a typical GMOS
field of view. Since this is about a factor of four larger than can be
accommodated in a single mask with GMOS, even in in Nod \& Shuffle
microslit mode, it is impossible to target every candidate $1<z<2$
galaxy with a single mask. On the other hand, to reach our required
depth on Gemini requires around 100ks of integration time, which means
it is not practical to obtain many masks per field in a single
semester.  Fortunately, the areal density of red galaxies with
$I-Ks>4$ and $I<24.5$ is only $\sim 1$ arcmin$^{-2}$, so it is at
least possible (in principle) to target all {\em red} galaxies in the
appropriate redshift range. We therefore adopted a sparse-sampling
strategy based on color, apparent magnitude, and photometric redshift
in order to maximize the number of targeted galaxies occurring in our
desired redshift range\footnote{In GDDS-SA02 and GDDS-SA22 the full
  color set was not available at the time of the mask design and so
  only $VRIz'K_s$ and $VIz'HK_s$, respectively, were used in these two
  fields. The impact of the smaller filter set for these two fields on
  the final sample selection was minor, as determined from tests with
  the GDDS-SA12 and GDDS-SA15 catalogs.}, with a particular emphasis
on red galaxies.  Targets were selected from the photometric catalogs
on the basis of $K_s$ and $I$ magnitudes measured in $3''$ diameter
apertures. Complete photometric catalogs for our sample will be
presented below.

\begin{figure*}[htbp]
\begin{center}
  \includegraphics[width=6.5in]{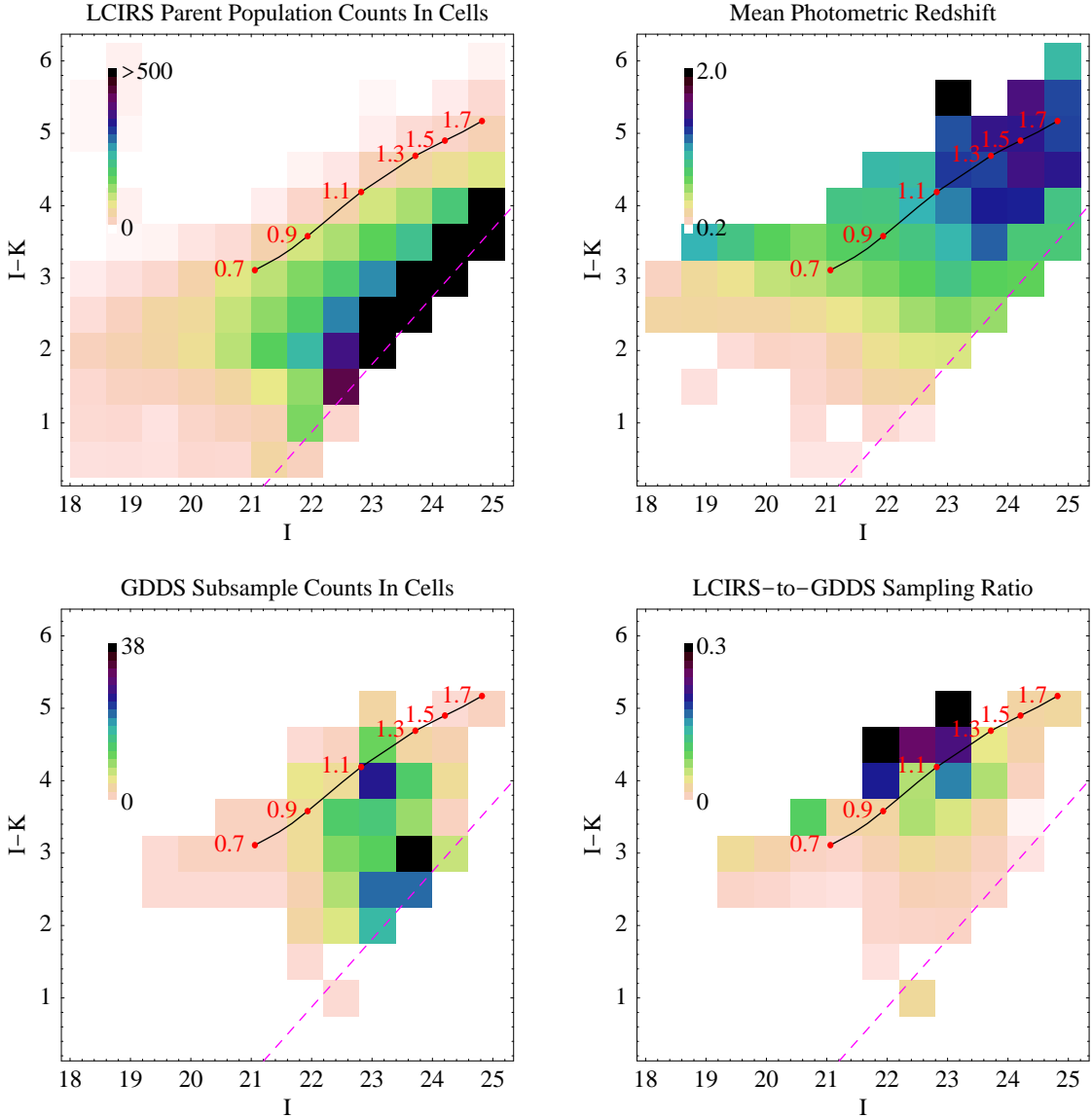}
\caption{\label{fig:lcirs} (Top left:) A two-dimensional histogram
  showing the distribution in color-magnitude space of all galaxies in
  a 554.7 square arcmin subset of the LCIRS, corresponding to the
  parent `tiles' of the LCIRS from within which our GDDS fields were
  chosen. The labeled track on this panel shows the expected position
  as a function of redshift of an M$^\star$ (assumed to be
  $M_K=-23.6$) galaxy formed in a 1 Gyr burst at redshift $z=10$. The
  position of this model old galaxy at several observed redshifts
  between $z=1.7$ and $z=0.7$ are marked with red dots and are
  labeled. (Top Right:) The photometric redshift in the LCIRS parent
  survey, computed from seven-filter photometry. (Bottom left:) A
  two-dimensional histogram showing the number of independent slits
  assigned each cell of color-magnitude space. (Bottom right:) The
  relative number of slits as a function of the average population in
  each cell expected in a wide-area survey, computed by dividing the
  bottom-left panel of the figure by the top-left panel.  These
  sampling weights are tabulated for each galaxy in the master data
  table accompanying this paper.  Note that dashed lines in each of
  these panels correspond to the detection limit of the survey. See
  text for additional details.}
\end{center}
\end{figure*}

Our sparse-sampling strategy is summarized graphically in
Figure~\ref{fig:lcirs}.  Each panel in this figure shows two-dimensional
histograms of different quantities within the parameter space defined by
our $(I-K_s)$ vs. $I$ photometry. The dashed line at the bottom-right
corner of each panel denotes the region in this parameter space below
which
the $K_s$ band magnitudes of galaxies become fainter than the formal
5$\sigma$ $K_s=20.6$ magnitude limit of the survey. Non-detections have been 
placed at the detection limit in our master data tables (presented below), so the 
bunching up of galaxies in the boxes intersected
by this line is mostly artificial (the counts are inflated by blue galaxies 
undetected in $K_s$).

The top-left panel of Figure~\ref{fig:lcirs} shows the distribution in
color-magnitude space of all galaxies in a 554.7 square arcmin subset
of the LCIRS, corresponding to the parent `tiles' of the LCIRS from
within which our GDDS fields were chosen. The labeled track on this
panel shows the expected position as a function of redshift of an
M$^\star$ (assumed to be $M_K=-23.6$) galaxy formed in a 1 Gyr burst
at redshift $z=10$.  The position of this model old galaxy at several
observed redshifts between $z=1.7$ and $z=0.7$ are marked with red
dots and are labeled.  Note the good agreement between the locus
defined by the reddest galaxies in the LCIRS and the model track. Most
galaxies should be bluer than our extreme old galaxy model, suggesting
that the optimal area for a mass-limited survey targeting $0.8<z<1.8$
should be the region defined by $\{22<I<24.5,\ 3<(I-K_s)<5\}$. This
basic conclusion is borne out by a comparison with the full
photometric redshift distribution computed from our seven-filter
photometry, shown in the top-right panel of Figure~\ref{fig:lcirs}.

Our strategy for assigning slits to objects was therefore based on the
following algorithm. Firstly, we assigned as many slits as possible to
objects with firm $K_s$ detections ($K_s < 20.6$ mag), red $I-K_s$
colors ($(I - K_s) > 3.5$ mag) and photometric redshifts greater than
0.8. As seen in Figure~\ref{fig:lcirs}, the red $(I-K_s)$ color
criterion alone gives a strong, but not perfect, selection against
redshifts below 0.8. Once the number of allocatable slits assigned to
such objects was exhausted, we then assigned slits to objects with
bluer $(I-K_s)$ colors but with firm $K_s$ detections and photometric
redshifts beyond $z=0.7$.  After doing this we started filling empty
space on the masks with objects whose $K_s$ photometry fell below our
$K_s$ detection limit, but who's photometric redshifts were greater
than 0.7. Such objects are, as a rule, the easiest to get redshifts
for but are our lowest priority, simply because our primary focus is
to learn more about the high-mass systems likely to be missed by other
surveys.

Our masks were designed to maximize the number of spectra per field.
With this goal in mind we used two tiers of low-dispersion spectra in
our mask designs, and laid out our slits so as to use the `microslit'
Nod \& Shuffle configuration described in Abraham et al. (2003) and in
Appendix~\ref{sec:nodandshuffle}. Each slit was 2.2\arcsec~ long
(corresponding to two 1.1\arcsec~long pseudo-slits at the two Nod \&
Shuffle positions) and 0.75\arcsec~wide (giving a spectral FWHM of
$\simeq 17$\AA)\footnote{The first mask (file GN2002BSV-78-14.fits in
  the Gemini archive) of our first field (GDDS-SA22) has
  2.0\arcsec~long slits. The slit length was increased by 10\% after
  an initial inspection of the incoming data suggested that 1.0\arcsec
  pseudo-slits were slightly too short for the largest galaxies being
  targeted.}. Additional room for charge storage on the detector must
be allowed for in the mask design, so each slit has an effective
footprint of 4.4\arcsec~ on the CCDs. Our two-tier mask design
strategy allowed spectral orders to overlap in some cases. A
classification system for describing these overlaps is presented in
Table~\ref{tab:collision}. Each two-dimensional spectrum has been
classified on this system and the results of this inspection are
included in the master data table presented in this paper. Our
strategy for dealing with order overlaps is described in detail in
Appendix C, but it is worth noting here that only modest overlaps were
allowed, and effort was made to ensure that top-priority objects
suffered little or no overlap.

\begin{deluxetable}{ll} 
\tablecolumns{2} 
\tablewidth{0pc} 
\tabletypesize{\small} 
\tablecaption{Classification of Spectrum Overlaps\label{tab:overlaps}}
\tablehead{ 
  \colhead{Class} & 
  \colhead{Meaning} 
} 
\startdata
 0 & Both A \& B channels uncontaminated (at most very minor masking needed). \\
 1 & Single channel overlap. Offending channel not used (at most very minor masking needed). \\
 2 & A contaminating 0th-order line has been masked. Remaining continuum is trustworthy. \\ 
 3 & Two channel collision. Major masking used in extraction. Continuum in blue should not be trusted. \\
 4 & Two channel collision. Major masking used in extraction. Continuum in red should not be trusted. \\
 5 & Extreme measures needed to try to recover a spectrum. Continuum should not be trusted.\\
\enddata
\label{tab:collision}
\end{deluxetable}

The number of slits per mask (not counting alignment holes) ranged
from 59 to 83 (see Table 1).  The highest density masks had a high
proportion of low-priority (non $K_s$-detected) objects and a somewhat
greater degree of overlap. To achieve this high slit density, objects
were distributed preferentialy in two vertical bands to the left and
right of the field center. Few objects were selected near the center
of the field, as this reduced the multiplexing options for that
portion of the field. Six masks were used over the four fields
(GDDS-SA12 and GDDS-SA22 had two masks each). In total 398 target
slits were cut into six masks, 323 of which were unique (since in the
cases where two masks were used in the same field, most slits on the
second mask were duplicates targeting the same galaxy as on the first
mask. The second mask was used simply because we had time between
lunations to quickly determine preliminary redshifts and drop obvious
low-redshift contaminants and replace these with alternate targets).
Our spectroscopic completeness ({\em i.e.} our success rate in turning
slits into measured redshifts) was around 80\%, with considerable
variation with both color and apparent magnitude. A detailed
investigation of spectroscopic completeness will be given in
\S\ref{sec:completeness}.

The practical upshot of our general mask design strategy is
graphically summarized in the bottom left panel of
Figure~\ref{fig:lcirs}. This panel is a two-dimensional histogram
showing the number of independent slits assigned each cell of
color-magnitude space. For the reasons just described heavy emphasis
is given to the $\{22<I<24.5,\ 3<(I-K_s)<5\}$ region of
color-magnitude space. The relative number of slits as a function of
the average population in each cell expected in a wide-area survey can
be computed by dividing the bottom-left panel of the figure by the
top-left panel. The values computed using this procedure are shown in
the bottom-right panel, and correspond to {\em sampling weights}.
These weights will prove important in the computation of the
luminosity and mass functions in future papers in this series. The
sampling weight for each galaxy in the GDDS is given in Table~4 which
will be presented later in this paper.

\section{SPECTROSCOPY}

\subsection{Observations}

The spectroscopic data described in this paper were obtained using the
Frederick C. Gillett Telescope (Gemini North) between the months of
August 2002 and August 2003. Most of data were taken by Gemini
Observatory staff using the observatory's queue observing mode. Data
were obtained only under conditions when the seeing was $<0.85$
arcsec, the cloud cover was such that any loss of signal was $<30\%$
at all times while on target, and the moon was below the horizon.
Typical seeing was in the range 0.45--0.65 arcsec as measured in the
$R$-band.

A detailed description of the procedure used to obtain and reduce the
data in the GDDS is given in Appendix A, and will only be outlined
here.  Our Nod \& Shuffle observations switched between two sky
positions with a cycle time of 120s, i.e. we spent 60s on the first
position (with the object at the top of the slit) and 60s on second
position (with the object at the bottom of the slit and charge
shuffled down) in each cycle. Fifteen such cycles gave us an 1800s
on-source integration which was read out and stored in an individual
data frame. Sky subtraction is undertaken by simply shifting each 2D
image by a known number of pixels and subtracting the shifted image
from the original image. Multiple 1800s integrations were combined to
build up the total exposure times given Table~1. Between exposures we
dithered spatially by moving the detector and spectrally by moving the
grating. As described in the appendices, these multiple dither
positions allow the effect of charge traps and inter-CCD gaps to be
removed from the final stack. By stacking spectra in the observed
frame and analyzing strong sky lines (OI, OH) we measured typical sky
residuals of 0.05\% -- 0.1\%, even with $>30$ hour integrations.

\begin{figure*}[htbp]
\begin{center}
\includegraphics[height=6.5in]{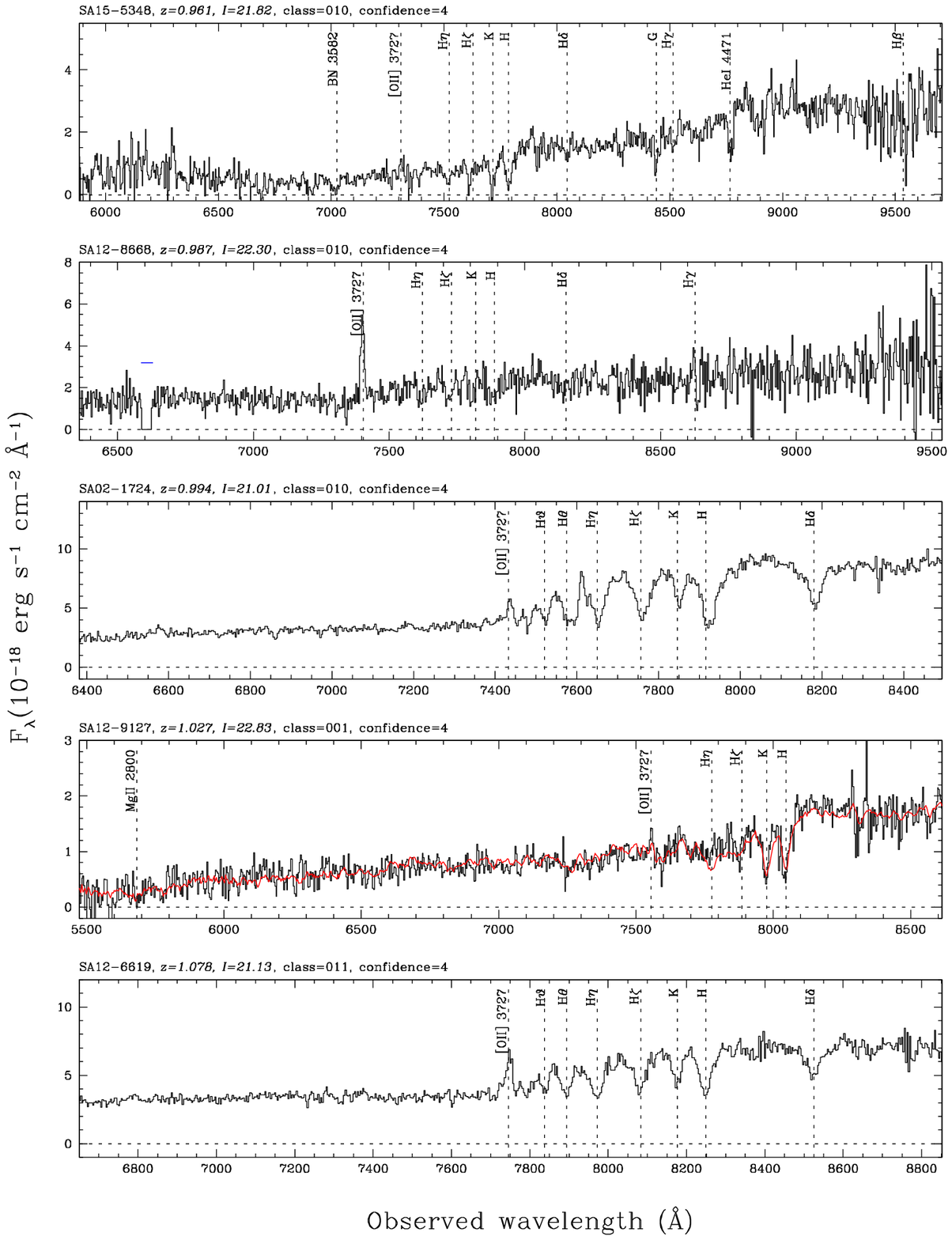}
\caption{
\label{fig:spectra1}
Representative GDDS spectra in the range $0.961<z<1.078$. Galaxies
with a range of spectral classifications are shown.  The SDSS luminous
red galaxy (LRG) composite (shown in red) has been overlaid on the
spectrum second from the bottom.  The LRG flux $F_\lambda$ is binned
to $\Delta \lambda = 2$ \AA\ rest frame, and rescaled to match the
observed spectrum according to the relation $a\times F_\lambda +b$
(where $a$ and $b$ are constants).  All galaxies shown in this figure
and the following four figures have redshift confidence classes of 3
or 4. See the text for further details.}
\end{center}
\end{figure*}

\begin{figure*}[htbp]
\begin{center}
\includegraphics[height=6.5in]{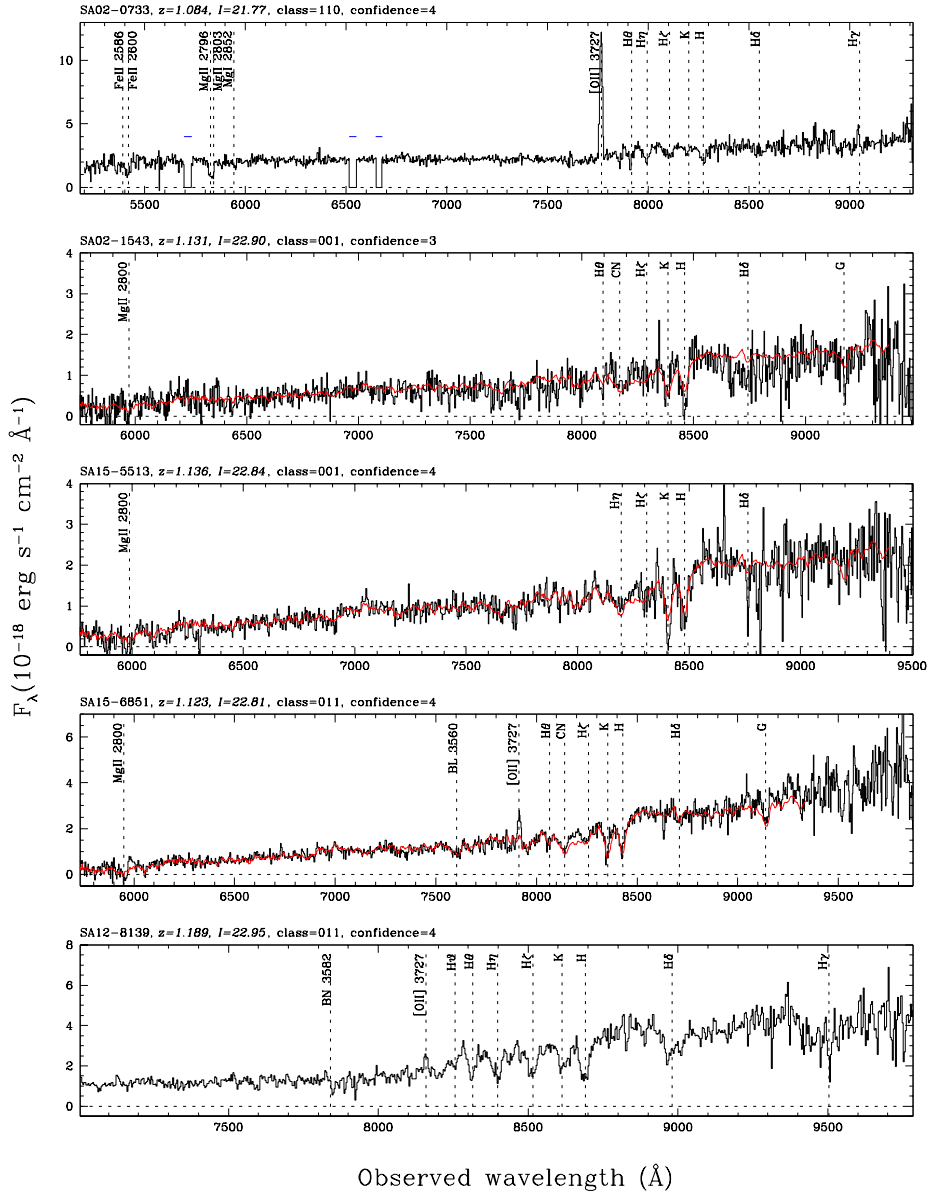}
\caption{
\label{fig:spectra2}
Additional GDDS representative spectra in the range $1.084<z<1.189$.
In cases where it is useful for identifying the redshift, the SDSS
luminous red galaxy composite is overlain on the spectrum (in red) as
in Fig. \ref{fig:spectra1}. Masked regions in spectra are indicated
using horizontal bars. See the caption of Figure~\ref{fig:spectra1}
for details.  }
\end{center}
\end{figure*}

\begin{figure*}[htbp]
\begin{center}
\includegraphics[height=6.5in]{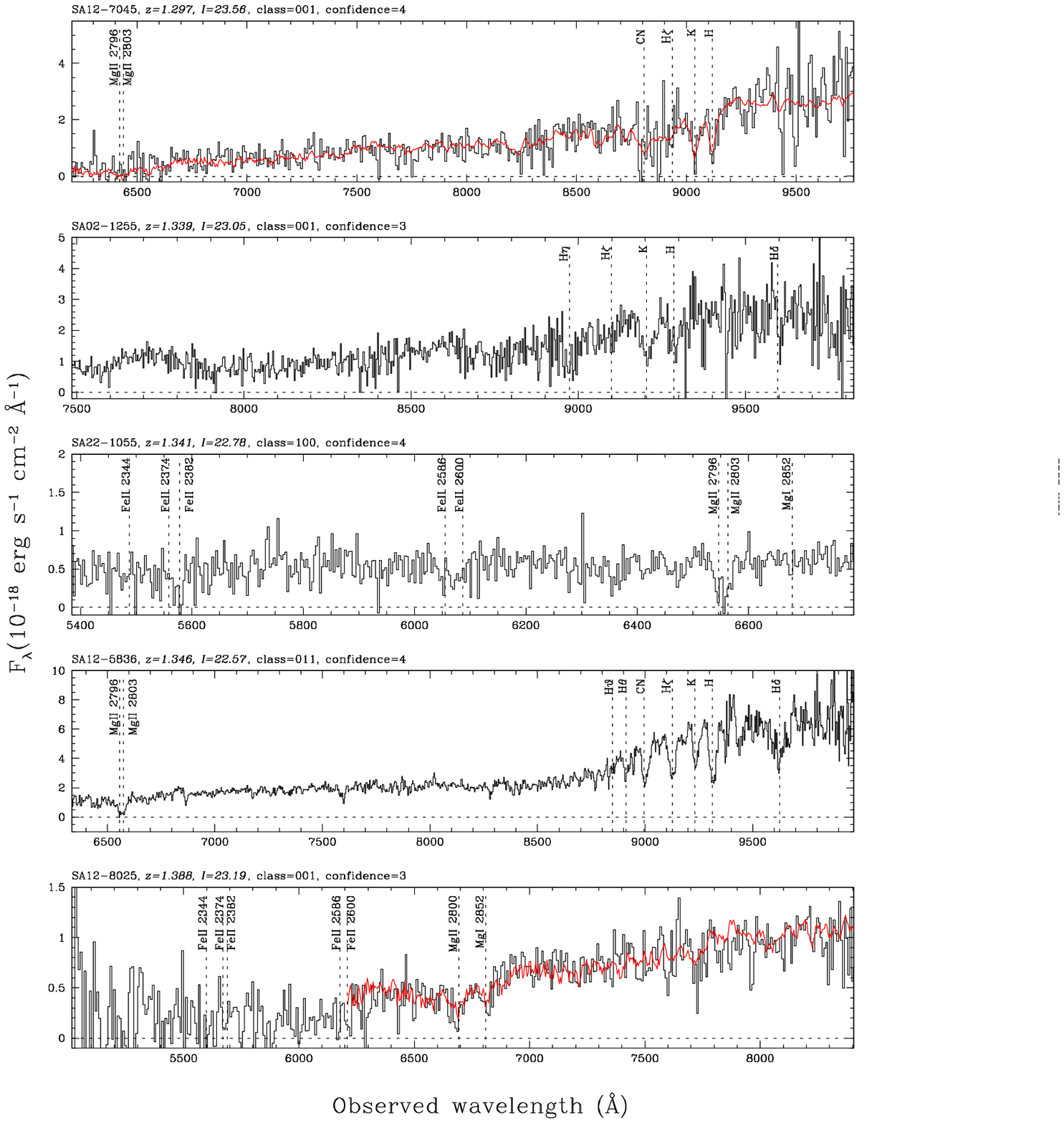}
\caption{
\label{fig:spectra3}
Representative GDDS spectra in the range $1.297<z<1.388$.  In cases
where it is useful for identifying the redshift, the SDSS luminous red
galaxy composite is overlain on the spectrum (in red) as in Fig.
\ref{fig:spectra1}. Masked regions in spectra are indicated using
horizontal bars. See the caption of Figure~\ref{fig:spectra1} for
details. The spectrum of SA12-7045 and SA12-8025 have been binned to a
pixel size of 8 \AA.}
\end{center}
\end{figure*}

\begin{figure*}[htbp]
\begin{center}
\includegraphics[height=6.5in]{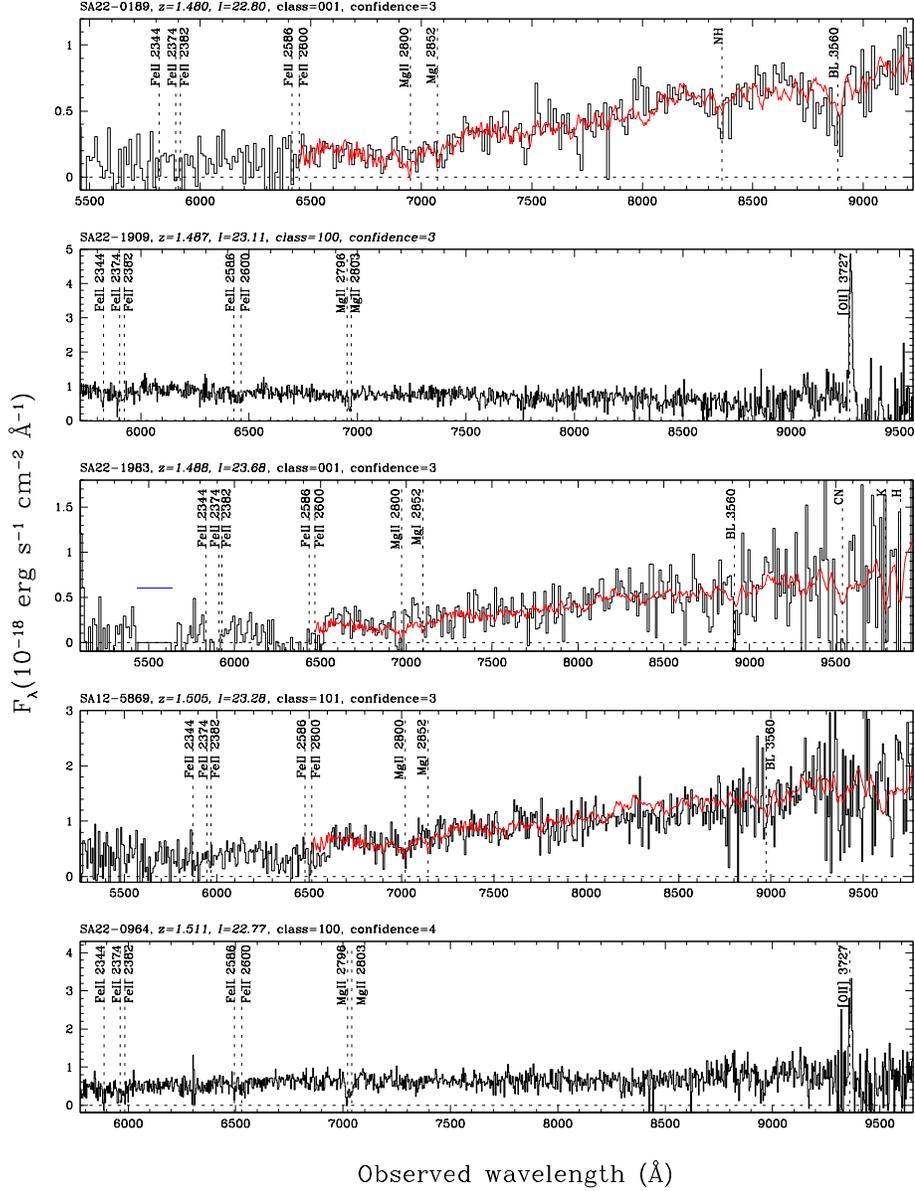}
\caption{
\label{fig:spectra4}
Representative GDDS spectra in the range $1.480<z<1.511$.  In cases
where it is useful for identifying the redshift, the SDSS luminous red
galaxy composite is overlain on the spectrum (in red) as in Fig.
\ref{fig:spectra1}. Masked regions in spectra are indicated using
horizontal bars. See caption of Figure~\ref{fig:spectra1} for details.
The spectrum of SA22-0189, SA22-1983, and SA12-5869 have been binned
to a pixel size of 12, 8, and 7 \AA, respectively.}
\end{center}
\end{figure*}

\begin{figure*}[htbp]
\begin{center}
\includegraphics[height=6.5in]{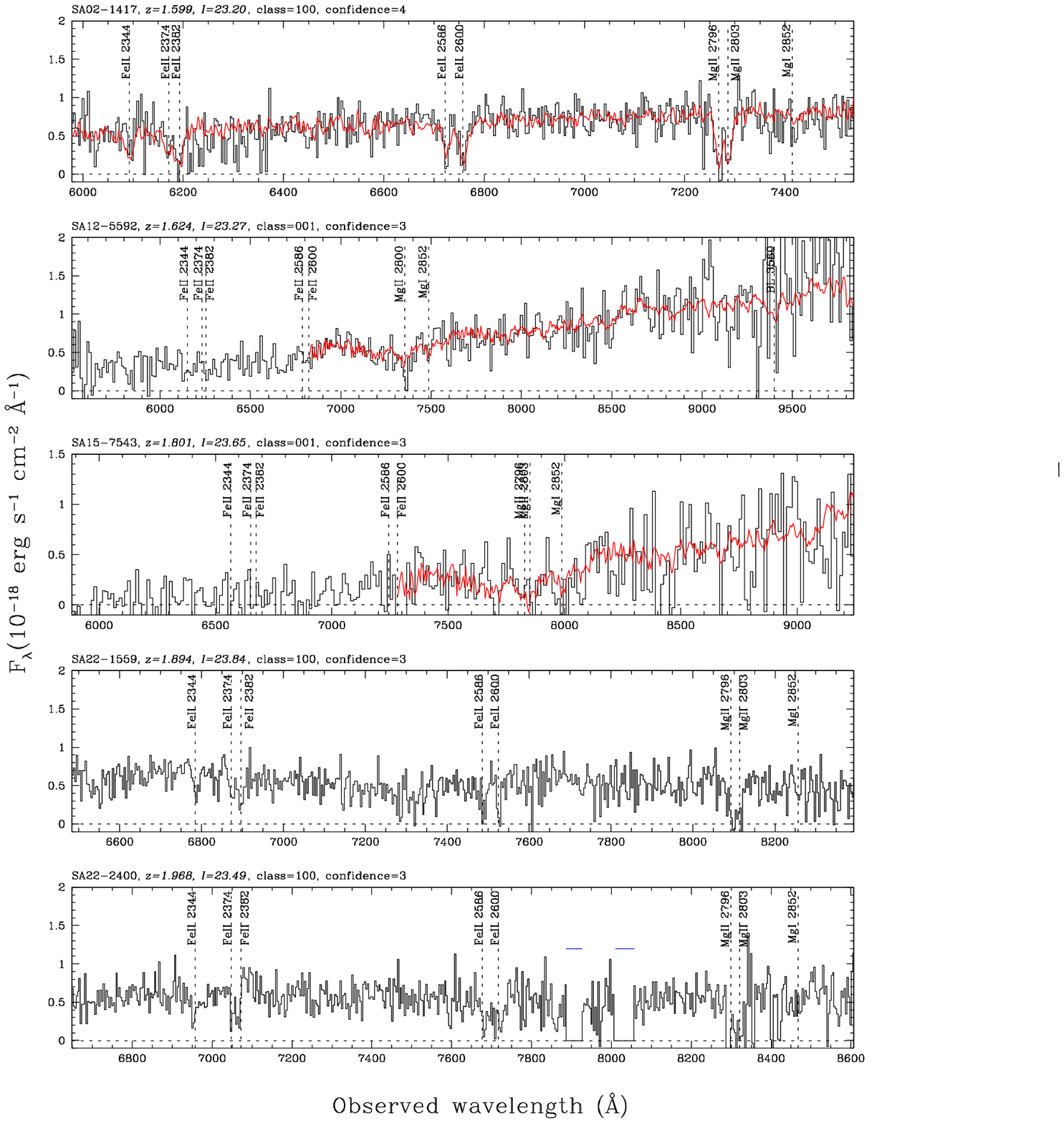}
\caption{
\label{fig:spectra5}
Representative GDDS spectra in the range $1.599<z<1.968$. In the top
panel a starburst composite kindly supplied by C. Tremonti is overlaid
on the GDDS spectrum. The SDSS luminous red galaxy composite is
overlaid on spectra second and third from the top (in red) as in Fig.
\ref{fig:spectra1}. Masked regions in spectra are indicated using
horizontal bars. See caption of Figure~\ref{fig:spectra1} for details.
The spectrum of SA12-5592 and SA15-7543 have been binned to a pixel
size of 12 \AA.}
\end{center}
\end{figure*}

The GMOS instruments are imaging multi-slit spectrographs capable of
spectroscopic resolutions ($\lambda/\Delta\lambda$) between 600 and
4000 over a wavelength range of 0.36--1.0\micron. Different detectors
are used on the Northern and Southern versions of the instrument. On
Gemini North the GMOS focal plane is covered by three 2048$\times$4608
EEV detectors which give a plate scale of 0.0727 arcsec per unbinned
pixel.  The GDDS observations were all taken with the R150\_G5306
grating in first order and the OG515\_G0306 blocking filter, giving a
typical wavelength range of 5500--9200\AA.  (In any multi-slit
configuration the exact range depends on the geometric constraints of
each individual slit.) This configuration gives a dispersion of
1.7\AA\ per unbinned pixel. Due to the high sampling of the CCD, all
data were taken binned by a factor of two in the dispersion direction
in order to reduce data volume and speed up readout, giving a final
dispersion of 3.4\AA per binned pixel.  Representative spectra from
the survey are shown in
Figures~\ref{fig:spectra1}--\ref{fig:spectra5}.

All of the multiple 1800s exposures (typically 30--60 spread across
multiple nights) making up a mask observation were sky-subtracted
(using the shuffled image) and combined into a single 2D frame called
the `supercombine'. Full details are given in Appendix B.  These were
then extracted to 1D spectra using standard procedures, but using
special software to allow efficient interactive assessment and
adjustment (see Appendix C for more details).

\subsection{Determination of Redshifts}

The absence of artifacts from poor sky subtraction made redshift
determination straightforward in the majority of cases. The reality of
weak features was also aided by the fact that the {\tt iGDDS} software
used for our analysis (described in Appendix~\ref{appendix:oned})
provides both one-dimensional and two-dimensional displays of the
spectra. An advantage of the Nod and Shuffle technique is that both
negative and positive versions of the two-dimensional spectra are
recorded and consequently real features display a distinctive pattern
that is easily recognized on the two-dimensional spectrum.

The presence of strong emission and absorption features (e.g., \OTwo,
\OThree, CaII, MgII, D4000, Hydrogren Balmer lines) immediately
indicated the approximate redshift in the majority of cases for
galaxies at $z<1.2$ At higher redshifts, [OII] can be seen in
star-forming objects to $z=1.7$, along with the blue UV continuum and
absorption lines (primarily MgII ad FeII). For redder objects, once
H\&K become undetectable we relied on template matching, which proved
to be an excellent aid to redshift estimation. Many such redshifts
were obtained using the interactive template manipulation tools built
into {\tt iGDDS}. Good templates covering the 2000-3500\AA~ wavelength
region for a variety of spectral types were not initially available
and we eventually constructed some of our own (from galaxies with
redshifts that were obvious from other features in their spectra).
These templates are shown in Section~\ref{sec:templates}, and proved
to be invaluable, particularly for spectra that just exhibit broad
absorption features and continuum shape variations in the observed
spectral range. On spectra for which the redshifts were uncertain or
indeterminate, cross-correlation versus a variety of templates were
used to suggest possible redshift/spectral feature matches.  These
templates included Lyman break galaxies \citep{sha01}, a composite
starburst galaxy template (kindly supplied by C. Tremonti), the red
galaxy composite from the Sloan Digital Sky Survey published in
\citet{eis03}, and a selection of nearby galaxy spectra obtained from
various sources (see Appendix~C). Ultimately the most useful templates
proved to be the ones constructed iteratively from the GDDS data
itself and presented in \S\ref{sec:templates}. Since the continuum
shape is an important indicator of galaxy redshifts but is not
utilized in the cross-correlation technique, all our final redshift
determinations were based on a combination of features, not just a
cross-correlation peak.

\begin{deluxetable}{ccl} 
\tablecolumns{3} 
\tablewidth{0pc} 
\tabletypesize{\small} 
\tablecaption{Redshift Confidence Classes\label{tab:confidence}}
\tablehead{ 
  \colhead{Class} & 
  \colhead{Confidence} &
  \colhead{Note}
} 
\startdata
\\
\cutinhead{\em Failures\hfill}
0 & None & No redshift determined. If a redshift is given in Table 4 it should be taken as an educated guess. \\
1 & $<50$\% & Very insecure.  \\ \\
\cutinhead{\em Redshifts Inferred from Multiple Features}
2 & $>75$\% & Reasonably secure. Two or more matching lines/features. \\
3 & 95\% & Secure. Two or more matching lines/features + supporting continuum. \\
4 & Certain & Unquestionably correct. \\ \\
\cutinhead{\em Single Line Redshifts}
8 & & Single emission line. Continuum suggests line is \OTwo. \\
9 & & Single emission line. \\ \\
\cutinhead{\em AGN Redshifts}
1$n$ & & Class $n$ as above, but with with AGN characteristics\\
\enddata
\end{deluxetable}

Each object's redshift was assigned a ``confidence class'', based on
the system adopted by Lilly et al. (1995) for the {\em Canada-France
  Redshift Survey}. This system is summarized in
Table~\ref{tab:confidence}. The confidence class reflects the
consensus probability (based on a quorum of at least five team
members) that the assigned redshift is correct, and takes into
consideration each spectrum's signal-to-noise ratio, number of
emission/absorption features, local continuum shape near prominent
lines (eg \OTwo), and global continuum shape. We did not factor galaxy
color into our redshift confidence classifications, which are
independent of photometric redshift, although a post-facto inspection
shows that the colors of essentially all single emission-line objects
(classes 8 and 9) are consistent with the line being \OTwo.  Redshift
measurements for the GDDS sample are presented in Table~4, along with
corresponding photometry for each galaxy. Note that in this table
non-detections have been placed at the formal 2$\sigma$ detection
limits and flagged with an magnitude error of $-9.99$. These detection
limits are $B = 27.5$ mag, $V = 27.5$ mag, $R = 27.0$ mag, $I = 25.5$
mag, $z = 24.5$ mag, $H = 21.0$ mag and $K_s = 21.0$ mag.


\subsection{Statistical Completeness}
\label{sec:completeness}

Our analysis of the statistical completeness of the GDDS is broken down
into two components. Firstly, there is the component of completeness
that quantifies the number of spectra obtained relative to the number of
galaxies that could possibly have been targeted. We refer to this
component of the completeness as the {\em sampling efficiency} of the
survey. Secondly there is the fraction of redshifts actually obtained
relative to the number of redshifts attempted. We will refer to this as
{\em spectroscopic completeness} of the survey.

\begin{figure*}[htbp]
\begin{center}
\includegraphics[width=5.5in]{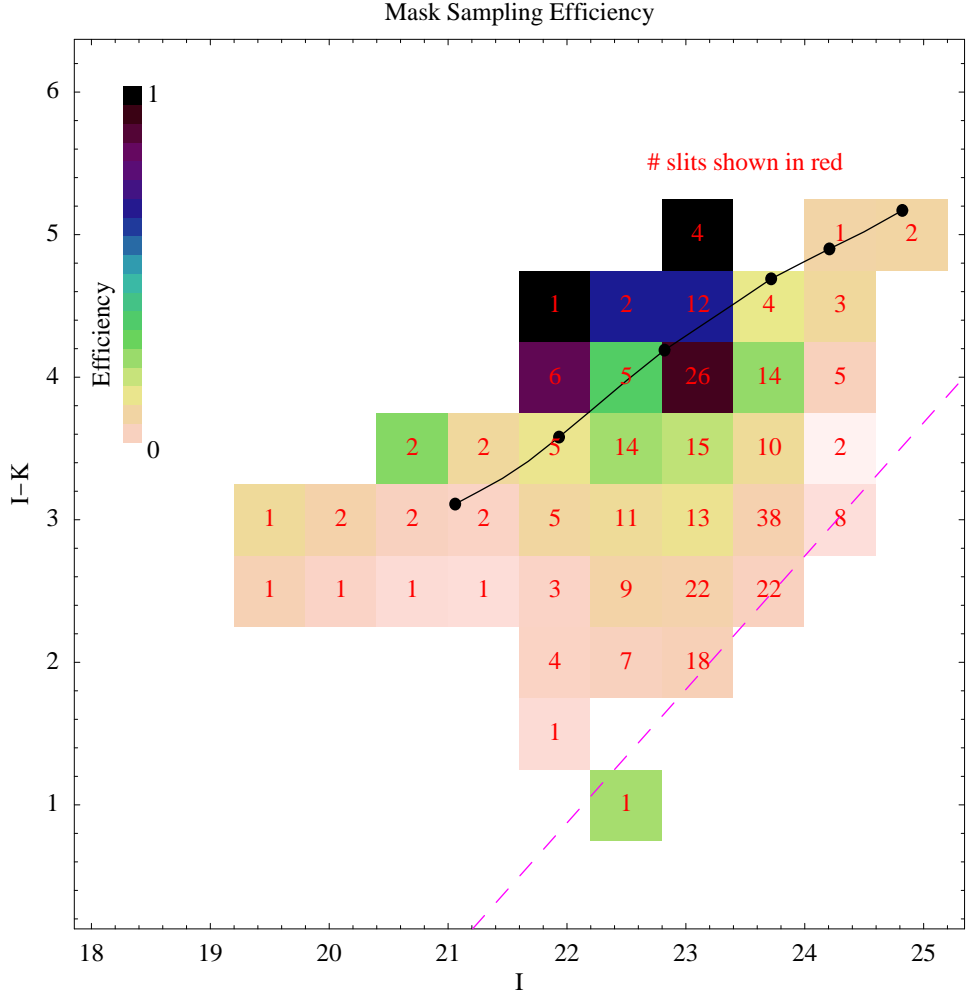} 
\caption{
\label{fig:samplingefficiency}
An analysis of sampling efficiency of our GDDS masks. As described in
the text, the number counts of faint galaxies at the magnitude limit
of the GDDS is much larger than can be targeted on a single
multi-object spectroscopic mask. The figure shows a two-dimensional
histogram with galaxy counts as a function of magnitude and color. The
color of each cell corresponds to the fraction of potential targets in
the GDDS images' field of view actually targeted with a slit.  Our
algorithm for assigning slits to objects is described in \S3 of the
text. The dashed line at the bottom-right corner denotes the region in
this parameter space below which the $K_s$ band magnitudes of galaxies
become fainter than the formal 5$\sigma$ $K_s=20.6$ magnitude limit of
the survey. As described in the text, the bunching up of galaxies in
the boxes intersected by this line is mostly artificial, since the
counts in these boxes are inflated by blue galaxies undetected in
$K_s$ that have been placed at the formal detection limit of the
survey in our data tables.}
\end{center} \end{figure*}

The sampling efficiency of the GDDS is investigated in
Figure~\ref{fig:samplingefficiency}.  This figure is essentially a
visual summary illustrating the success of the mask design algorithm
described in \S\ref{sec:design}. The figure shows a two-dimensional
histogram quantifying the number of slits assigned to targets as a
function of color and magnitude. The sampling efficiency of each cell
is keyed to the color bar also shown in the figure. As described
earlier, our mask design strategy places heavy emphasis on targeting
objects with colors and apparent magnitudes consistent with those of a
passively evolving luminous galaxy at $0.8<z<1.8$. Slits were placed
on essentially all red galaxies with apparent magnitudes around
$I=23.5$ mag (corresponding to the expected brightness of an $M^\star$
galaxy at $I\sim1.3$).  The number-weighted average of the sampling
efficiency over the cells in the optimal region of this diagram
(described \S\ref{sec:design}, and corresponding to $\{22<I<24.5,\
3<(I-K_s)<5\}$) is 50\%. Thus the GDDS can be thought of as a
one-in-two sparse sample of the reddest and most luminous galaxies
near the track mapped out by passively evolving high-redshift galaxies
in $I$ vs. $I-K$.  This sample is augmented by a one-in-seven sparse
sample of the remaining galaxy population.

\begin{figure*}[htbp]
\begin{center}
\includegraphics[width=5.5in]{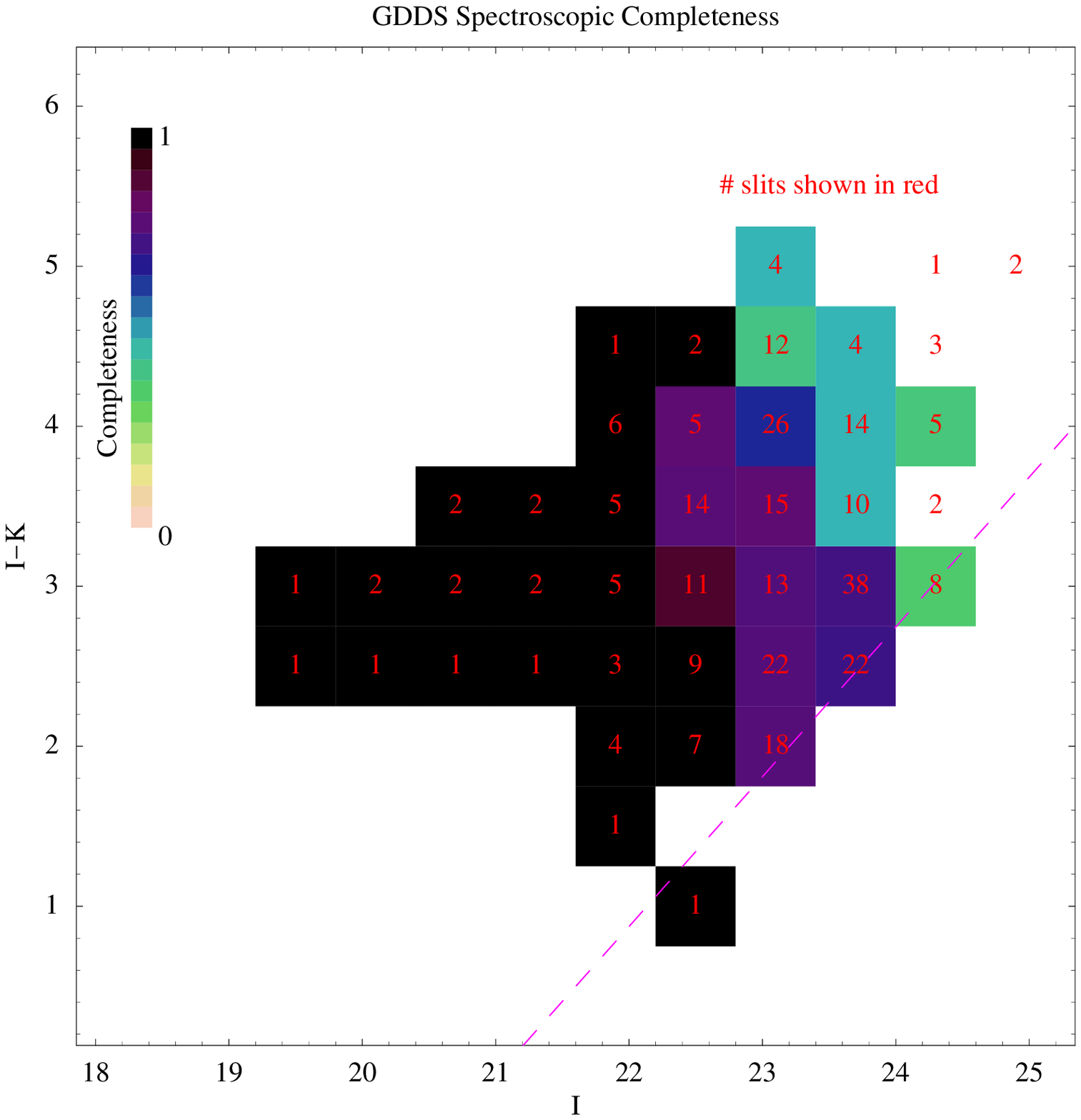} 
\caption{\label{fig:completeness1}
  An analysis of spectroscopic completeness in the GDDS fields. This
  two-dimensional histogram quantifies the success rate in recovering
  redshifts as a function of color and magnitude. (Note that only
  those objects with redshift confidence classes of three or higher
  were considered to be successfully measured). Red numbers embedded
  in each cell correspond to the number of slits assigned to that
  portion of the parameter space. The color of the cell denotes the
  spectroscopic completeness, keyed to the color bar at the top left.
  The dashed line at the bottom-right corner of the panel denotes the
  region in this parameter space below which the $K_s$ band magnitudes
  of galaxies become fainter than the formal 5$\sigma$ $K_s=20.6$
  magnitude limit of the survey. As described in the text, the
  bunching up of galaxies in the boxes intersected by this line is
  mostly artificial, since the counts in these boxes are inflated by
  blue galaxies undetected in $K_s$ that have been placed at the
  formal detection limit of the survey in our data tables.}
\end{center}
\end{figure*}

Our success in translating slits into redshifts is shown in
Figure~\ref{fig:completeness1}.  This diagram is the close analog to the
previous figure, with the difference being that cell colors are keyed to
spectroscopic completeness instead of sampling efficiency. Spectroscopic
completeness is calculated simply by dividing the number of high-confidence
redshifts (confidence class greater than or equal to 3) in each cell by the
number of slits assigned in the same cell.  As expected, spectroscopic
completeness is a strong function of apparent magnitude. Spectroscopic
completeness is 100\% brighter than $I=22$ mag, dropping to around 50\% at
$24.0<I<24.5$ mag (11 redshifts out of 20 attempted).

The overall spectroscopic completeness of the GDDS depends on the minimum value
of the redshift confidence that is considered acceptable. Slits were placed on
317 objects. Three of these slits contained two objects and thus redshift
determinations were attempted for 320 objects. Of these, approximately 3\% (11
objects) were so badly compromised by overlap or contamination that they were
judged to be invalid. Of the 309 valid spectra, approximately 79\% of the
attempts resulted in moderately high confidence (classes 2 and 9) or very high
confidence (classes 3,4, and 8) redshifts. As will be shown in
\S\ref{sec:ultimate}, the great majority of these were in our target redshift
range, with only a very modest contamination by very low redshift objects.
(Twelve objects were found to be late-type galactic stars, and 5 objects were
found to be $z \sim 0.1$ extragalactic HII regions). An additional 10\% of GDDS
targets had very low confidence redshifts assigned (class 1 and 0). The
remaining 11\% of targets had no redshifts assigned. 

\begin{figure*}[htbp]
\begin{center}
\includegraphics[width=6in]{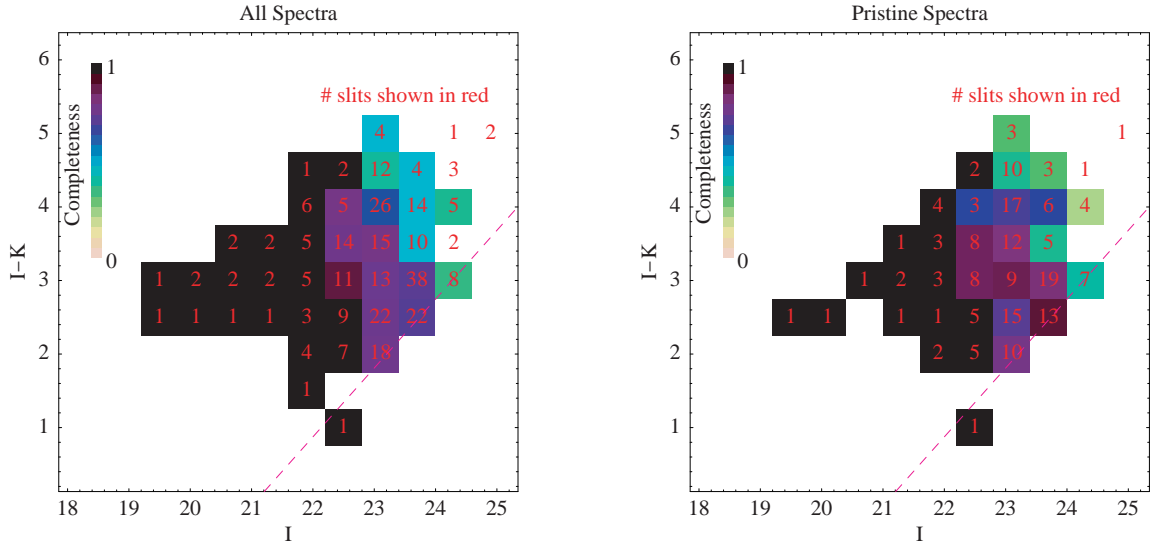} 
\caption{\label{fig:pristine}
  An analysis of the relative completeness for our full sample (shown
  at left), and a subsample constructed from only those spectra
  unaffected by spectrum overlaps (shown at right). The dashed line
  corresponds to the completeness limit of the survey.  }
\end{center}
\end{figure*}

\begin{figure*}[htbp]
\begin{center}
\includegraphics[width=6.5in]{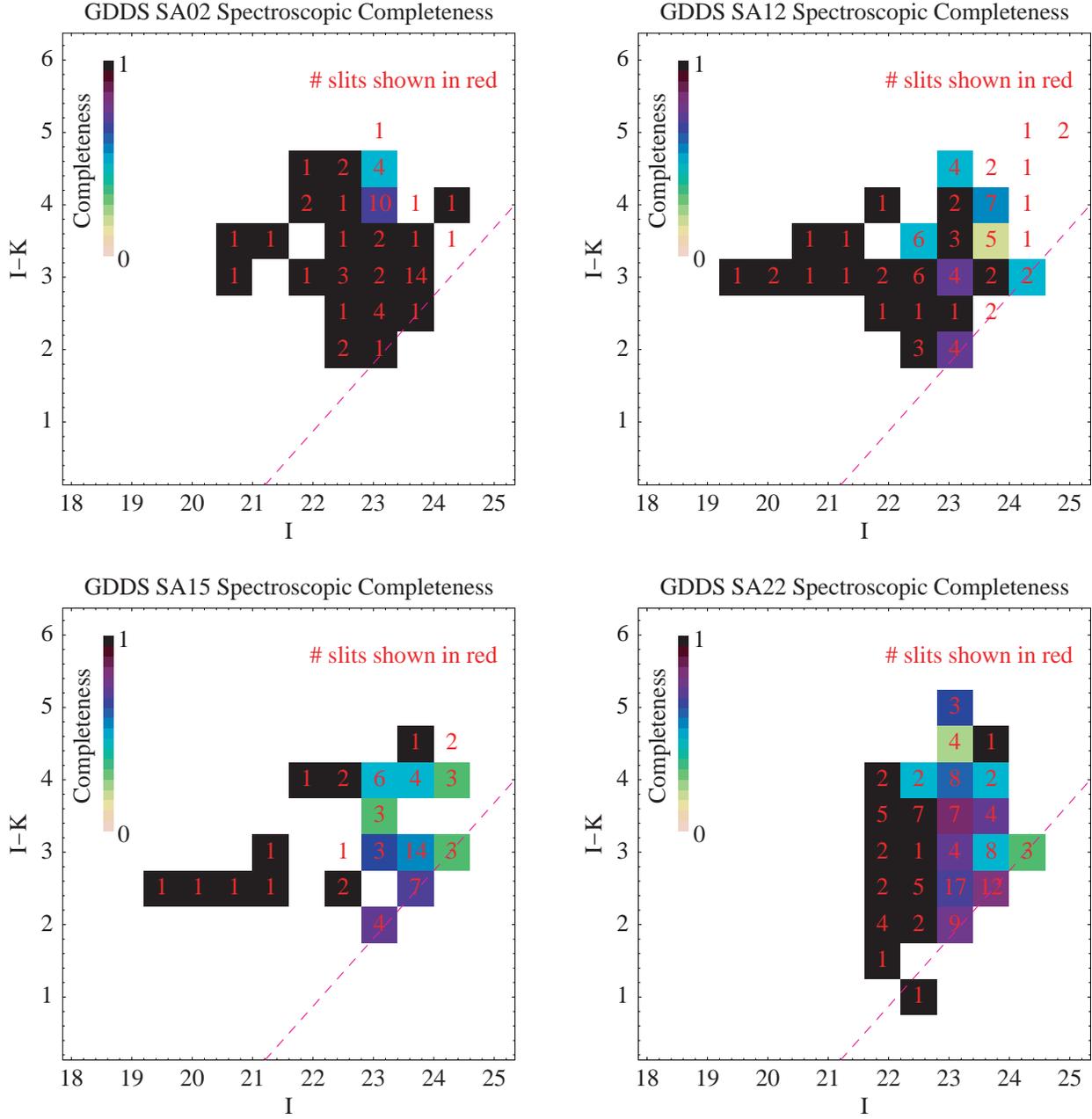} 
\caption{\label{fig:completenessFieldToField} An analysis of the
field-to-field variation in the spectroscopic completeness in the
GDDS. Two-dimensional completeness histograms for GDDS-SA02
(top-left), GDDS-SA12 (top-right), GDDS-SA15 (lower-left), and
GDDS-SA22 (lower-right) are shown. As for all similar figures in this
paper, the dashed line at the bottom-right of the figure denotes the
region of color-magnitude space below which measured $K_s$ photometry drops
below the formal magnitude limit of the survey. See the caption to the previous figure for
additional details.  }
\end{center}
\end{figure*}

It is interesting to consider the main sources of spectroscopic
incompleteness in the GDDS. The obvious gradient in incompleteness as
a function of apparent magnitude seen in
Figure~\ref{fig:completeness1} indicates that the main cause of
incompleteness is photon starvation, particularly for red systems with
little rest-UV flux. However, in some cases we were unable to obtain
redshifts for relatively bright systems on account of overlapping
spectral orders. Our mask design strategy was a trade-off between
maximizing the number of objects we attempted to get redshifts for,
and minimizing spectrum order overlaps. It is worthwhile to try to
quantify the relative importance of these competing effects. An
attempt at doing this is illustrated in Figure~\ref{fig:pristine},
which shows a comparison between the cell-to-cell completeness in our
full sample and the corresponding cell-to-cell completeness in a
sub-sample of ``pristine'' objects unaffected by spectrum overlaps.
Note that the spectroscopic completeness of our red galaxy sample is
nearly unchanged in both panels of this figure, while greater
variation is seen in the blue population.  As described in
\S\ref{sec:design}, our mask-design strategy was optimized for red
galaxies, and a particular effort was made to avoid spectrum overlaps
in this population in order to preserve continuum shape. On the other
hand, our emphasis in laying down slits on blue galaxies was to use
the detector area efficiently even at the expense of sometimes
allowing spectrum overlaps to occur (since emission line redshifts can
often be determined from these by studying the two-dimensional
spectra). We therefore expected to see a somewhat greater variation in
the cell-to-cell completeness of blue galaxies as a function of slit
overlap class, and Figure~\ref{fig:pristine} is consistent with this.

We also note that the extra freedom used when assigning slits to blue
galaxies results in substantially greater field-to-field variation in
the spectroscopic completeness of blue galaxies in the GDDS. The
field-to-field completeness of the GDDS is shown in
Figure~\ref{fig:completenessFieldToField}. Significant variations in
the total completeness in each field are seen, ranging from a high of
87\% in GDDS-SA02, to a low of 60\% in GDDS-SA12. It is important to
note that most of this field-to-field variation is in the blue
population, and the cell-to-cell completeness near the red galaxy
locus of the color-magnitude diagram in each panel of this figure
remains quite stable.

\subsection{Spectral Type Classification}

In addition to the redshift confidence class, each object was assigned
a series of spectral classifications that record both the features
that were used to determine the redshift and a subjective
classification of the galaxy's spectral type. Our spectral
classifications are presented in Table~5. The first column notes
whether any features indicative of AGN activity are seen in the
spectrum (0 = no, 1 = yes). The next 11 entries specify the presence
(1) or lack (0) of the most common spectral features used in the
redshift determinations. A (2) in any of these columns indicates that
the particular feature did not fall within the wavelength range of our
spectra. In some cases the spectral range covered by a particulary
object was reduced by overlap or contamination from other objects. The
``template" column in Table 5 identifies those objects whose redshifts
were based largely on a match to a template spectrum, either a
composite from our own spectra (see below) or an external spectrum. It
is emphasized that spectra on different masks had different exposure
times, and order overlaps impacted some spectra more than others,
so these feature visibility classifications should be used as a general
guideline only and not over-interpreted.

The last column in Table 5 lists the spectral class assigned to each
object.  This classification is based on three digits that flag
young, intermediate age, and old stellar populations.  Objects showing
pure, or nearly pure, signatures of an evolved stellar population
(e.g. D4000, H\&K, or template matches) are assigned a class of
``001". Objects that are dominated by the flat-UV continuum and strong
emission-lines characteristic of star forming systems are assigned a
``100" classification, those showing signatures of intermediate ages
(e.g. strong Balmer absorption) are assigned a class of ``010". Many
objects show characteristics of more than one type and so are assigned
classes that are the composite of old (001), intermediates (010) and young
(100) populations. Objects listed as ``101" may show strong H\&K
absorption and 4000\AA~ breaks and yet have a flat-UV continuum tail
indicative of a low-level of ongoing star formation.

\section{SUMMARY OF REDSHIFTS AND SPECTRAL CLASSIFICATIONS}
\label{sec:ultimate}

\begin{figure*}[htbp]
\begin{center}
\includegraphics[height=5.6in,angle=90]{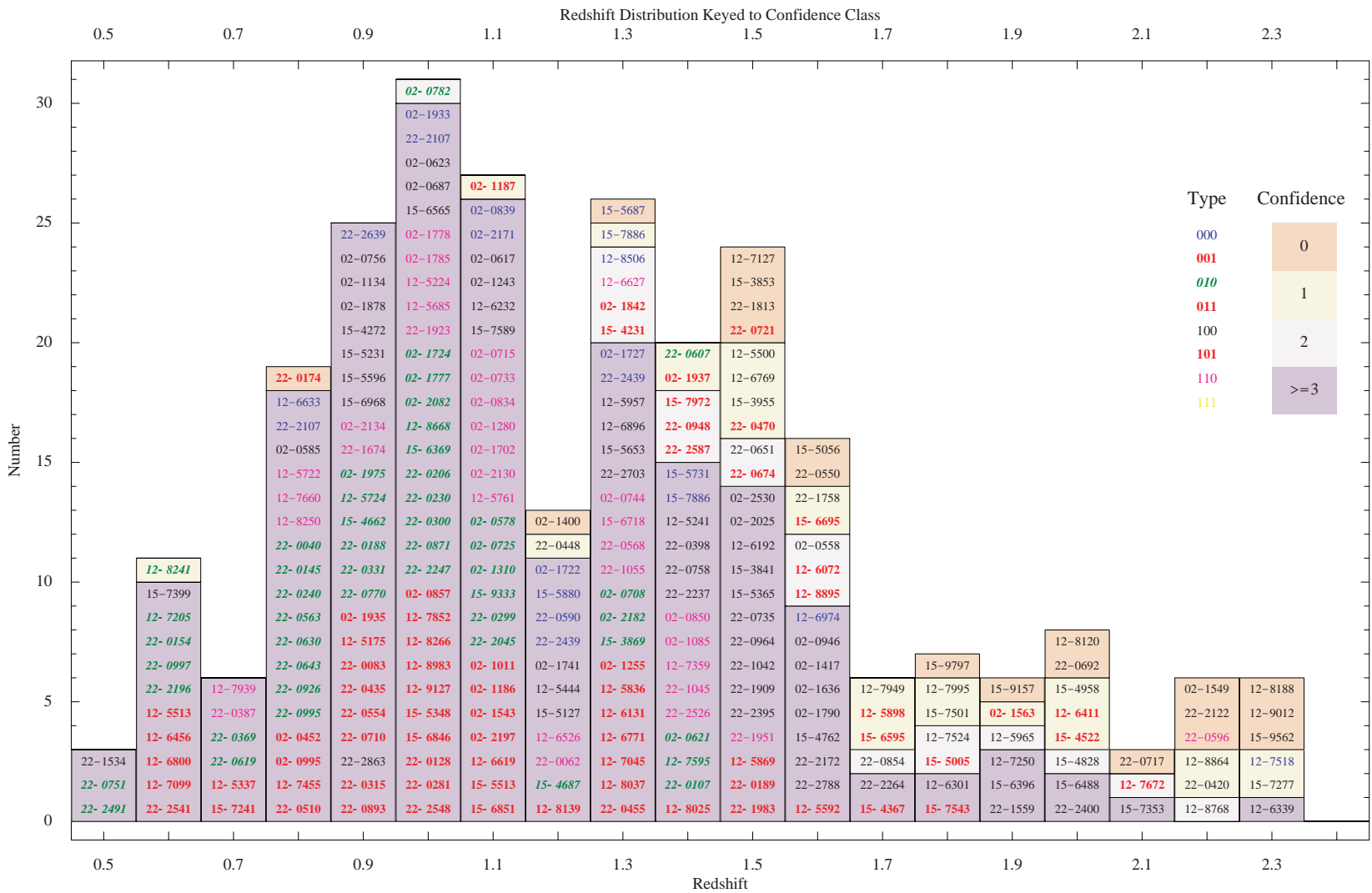} 
\caption{\label{fig:ultimate}
  Master summary of the GDDS. This histogram illustrates the redshift
  distribution of our sample as a function of confidence class and
  spectral type. Bar colors indicate confidence in the redshifts,
  keyed to the system defined in Table~2. Individual objects are
  labeled in each histogram and sorted by spectral class. Label colors
  are keyed to spectral class.  }
\end{center}
\end{figure*}

A graphical synopsis of the spectroscopic information in the GDDS is
presented in Figure~\ref{fig:ultimate}, which shows the number vs.
redshift histogram for our sample, color-coded both by confidence
class and by spectral class. The shading of the box shows the
confidence class, the color of the label reflects the spectral class.
A number of interesting aspects of our experiment are evident in
Figure~\ref{fig:ultimate}. Firstly, despite our use of four widely
separated fields, we remain significantly impacted by large-scale
structure and sample variance. The factor of two deficit of objects at
$z = 1.2$ cannot be the sign of failure to recognize objects at this
redshift as we do considerably better at $z = 1.3$, a more difficult
redshift. Secondly, it is clear that our success rate drops steeply at
$z > 1.5$, where our fraction of high confidence redshifts drops from
$> 90$\% to $\sim 30$\%. As described in \S\ref{sec:completeness},
this is mostly because these objects, and the red ones in particular,
are fainter than the average galaxy in the sample, particularly at
wavelengths shortward of the $I$-band used to set the magnitude limit
of the sample.

The following summarizes the fractions of different stellar populations amongst objects with high-confidence redshifts in the GDDS. Approximately 15\% of the objects observed showed spectra with pure old stellar populations (class 001). The fraction with strong signatures of evolved stars (i.e. objects with classes 001 or 011) is 22\%.  Galaxies showing some evidence for intermediate age stellar population features (classes 110, 010, 001) accounted for 46\% of the sample. About 25\% of galaxies had pure intermediate age populations (class 010), and 24\% of the sample appear to be dominated by young populations (class 100). We were unable to assign any spectral classifications to 10\% of objects with high confidence redshifts.

The fraction of galaxies with evidence of old populations peaks at $z
\sim 1$ and falls off steeply at higher redshifts. Some of this
reflects the increasing impact of the non-$K_s$-detected objects that
were added to the sample on the basis of their photometric redshifts.
A number of the $z > 1.5$ objects, however, do have $K < 20.6$ and
some of these have red $I - K_s$ colors and yet still show essentially
flat UV spectra dominated by massive stars. This is not surprising
given the shape of the $V-I$ vs. $I-H$ (or $I-K_s$) two-color diagram
(McCarthy et al. 2001). At $z > 1.5$ the bulk of the red $I-K_s$
population has blue $V-I$ colors. The spectroscopic properties of
these galaxies, along with inferred ages from stellar population
synthesis models, will be presented in a companion paper (McCarthy et
al. 2004, in preparation).

\section{COMPOSITE TEMPLATE SPECTRA}
\label{sec:templates} 

As described earlier, redshifts were determined by visual examination
of the spectra, comparing with spectral templates and looking for
expected redshifted emission and absorption features. The most
uncertain aspect at the start of our survey was the appearance of
normal galaxies in the 2000--3000\AA\ region.  Early-type quiescent
galaxy spectra are expected to be dominated by multiple broad
absorption features in this region (for example see the mid-UV spectra
of elliptical galaxies in Lotz et al. 2000) which come primarily from
F \& G main sequence stars (Nolan, Dunlop \& Jimenez 2001). In
contrast late-type actively star-forming galaxies should have a
featureless blue continuum with narrow ISM absorption lines
superimposed (for example see the HST starburst spectra in Tremonti et
al. 2003).

\begin{figure*}[htbp]
\begin{center}
\includegraphics[width=4.5in,angle=-90]{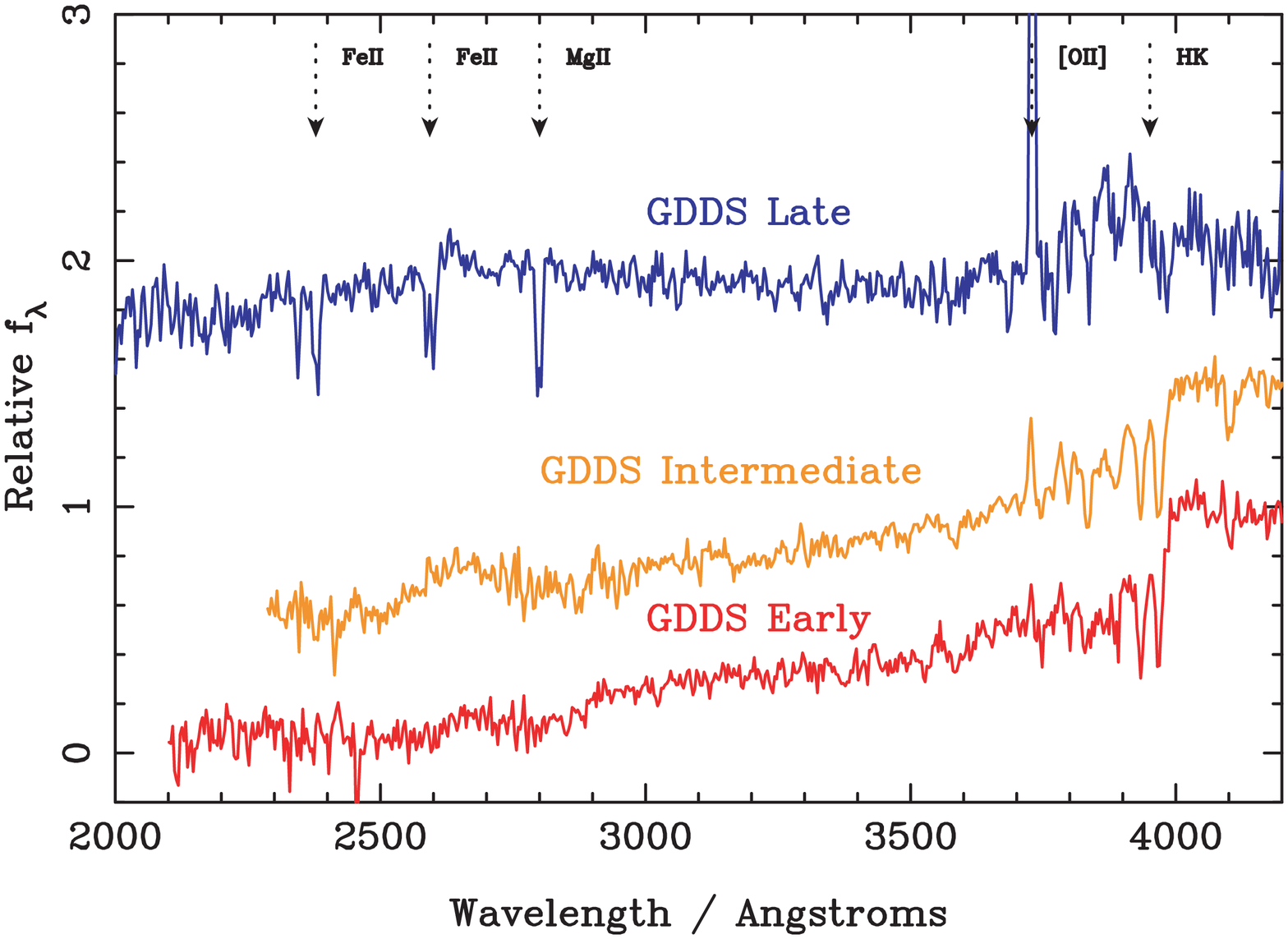} 
\caption{The three GDDS template composite spectra used for redshift
  comparison.  These were constructed internally from our sample from
  galaxies covering a range $0.6<z<2$ with high confidence redshifts.
  The templates are plotted at 3.5\AA\ per pixel resolution and are
  normalized to unity at 4200\AA, they are offset vertically in the
  plot for clarity. Some of the more common features are marked.  Only
  the UV portion of the templates is shown.}
\label{fig:templates}
\end{center}
\end{figure*}

For the first two GDDS fields we primarily used the Luminous Red
Galaxy template of \citet{eis03} and spectra of the $z\sim 1.4$ radio
galaxies 53W091 (Spinrad et al., 1997) and 53W069 (Dunlop 1999) for
our early-type reference templates. For late types we used a composite
spectrum made from an average of the local starburst galaxy
starforming regions of Tremonti et al. After we had reduced our first
two fields and obtained preliminary redshifts we then constructed our
own templates by combining GDDS spectra. We had three principal
motivations for doing this. The first was to obtain a better match in
spectral resolution; the second was to improve the UV coverage
especially in the early-type template and the third was to make
templates corresponding to galaxies in an earlier evolutionary epoch
in the history of the Universe.

In order to make the templates we visually identified similar looking
spectra with confidence level 3 or 4 redshifts. The full redshift
range of GDDS $0.6<z<2$ was used in order to produce templates with a
wide range of spectral coverage. Of course this large redshift range
is less than ideal because on average the resulting UV portion of the
templates will come from higher redshift objects than the optical
portion. However we decided that since the primary use of these
templates were for redshift matching, the wider wavelength coverage
would be of greater importance. It remains true that the UV portion of
the templates ($\lambda<3000$\AA) comes mostly from $z>1$ objects and
this was where our previous set of templates were most inadequate. We
emphasize that these redshift templates are {\em different\/} from the
composite spectrum analyzed by Savaglio et al. (2004), where it is
more important to have a more restricted redshift range.

The template construction process fully allowed for masking and
disparate wavelength coverage in the different spectra. Given a set of
input spectra the construction process was as follows. Firstly one
spectrum (typically at the median redshift) was chosen as a master for
the purposes of normalization. Next all the other templates were
scaled to the same normalization as the master by computing the
average flux in the non-masked overlapping regions. Finally a masked
average was performed of all the normalized templates (the mask being
either 1 or 0 depending on whether a given spectrum included that part
of the wavelength region with good data).

For the early type template we combined 8 convincing early type
spectra with redshifts $0.6<z<1.5$ and for the late type template we
combined 23 late type spectra with $0.8<z<2.0$. We also found it
desirable to make an intermediate type template (i.e., somewhat red but
with signs of star-formation such as [OII] emission) as these were
especially poorly represented in our first template set.  For this we
used 8 galaxies with $0.7<z<1.3$. The resulting three templates are
plotted in Figure~\ref{fig:templates}.

\section{CONCLUSIONS}
\label{sec:summary} 

The Gemini Deep Deep Survey was undertaken in order to explore galaxy
evolution near the peak epoch of galaxy building. The survey probes a
color-selected sample in a manner that minimizes the strong
star-formation rate selection biases inherent in most high-redshift
galaxy surveys. It is designed to bridge the gap between landmark
surveys of highly complete samples at $z<1$ \citep{cow94,lil95,ell96}
and UV-selected surveys at higher redshift \citep{ste96,sha01,ste03}.
The signal-to-noise ratios of the spectra in the GDDS are sufficient
to distinguish old stellar populations (dominated by F-type stars)
from post-starburst systems (dominated by A-type stars) and reddened
starbursts with their flat spectra and strong interstellar lines. In
this paper we have described the motivation for the survey, the choice
of fields, the experimental design underlying our choice of targets,
and our data reduction process. We have presented final catalogs of
redshifts and photometry and an analysis of their statistical
completeness. Spectra for individual objects are available as an
electronic supplement to this paper. Further information on the GDDS
and the data reduction software used in the project are available on
the World Wide Web at \url{http://www.ociw.edu/lcirs/gdds.html}.

\acknowledgments
\centerline{\em Acknowledgments}

\vskip 0.5 cm

\noindent 

This paper is based on observations obtained at the Gemini
Observatory, which is operated by the Association of Universities for
Research in Astronomy, Inc., under a cooperative agreement with the
NSF on behalf of the Gemini partnership: the National Science
Foundation (United States), the Particle Physics and Astronomy
Research Council (United Kingdom), the National Research Council
(Canada), CONICYT (Chile), the Australian Research Council
(Australia), CNPq (Brazil) and CONICET (Argentina)

The Gemini Deep Deep survey is the product of a
university/institutional partnership and would not have been possible
without the work of many people. We thank Matt Mountain, Jean-Ren\'e
Roy and Doug Simons at the Gemini Observatory for their vision in
supporting this project in the midst of many other observatory
priorities, and for the time, energy and manpower they have invested
in the making Nod \& Shuffle a reality on Gemini. It is a pleasure to
thank Matthieu Bec and Tatiana Paz at the Gemini Observatory for their
work in implementing the modifications to the GMOS telescope control
system and sequence executor needed in order to support the Nod \&
Shuffle mode.  We are grateful to the entire staff of the Gemini
Observatory for undertaking the queue observing for this project in
such an efficient manner, and for the kind hospitality shown to us
during our visits.

We also thank the Instrumentation Group at the Herzberg Institute of
Astrophysics for working with us and with Gemini in order to help make
GMOS Nod \& Shuffle a reality.  Many members of the HIA
Instrumentation Group went beyond the call of duty in support of this
project, but we would like to particularly thank Bob Wooff and Brian
Leckie at HIA for their quick response and willingness to work late
into the night during a weekend to fix some early problems we
encountered, which saved us from losing a night of observing during
the science verification phase of the GDDS.  We also thank Richard
Wolff at NOAO for doing much of the microcode programming needed for
this project.

RGA and RGC acknowledge generous support from the Natural Sciences and
Engineering Research Council of Canada, and RGA thanks the Government
of Ontario for funding provided from a Premier's Research Excellence
Award.  KG \& SS acknowledge generous funding from the David and
Lucille Packard Foundation. H.-W.C. acknowledges support by NASA
through a Hubble Fellowship grant HF-01147.01A from the Space
Telescope Science Institute, which is operated by the Association of
Universities for Research in Astronomy, Incorporated, under NASA
contract NAS5-26555.

\bigskip
\appendix
\bigskip
\centerline{\Large\bf Appendices}

\section{NOD \& SHUFFLE OBSERVATIONS WITH GMOS}
\label{sec:nodandshuffle}

\begin{figure*}[htbp]
\begin{center}
\includegraphics[width=6.5in]{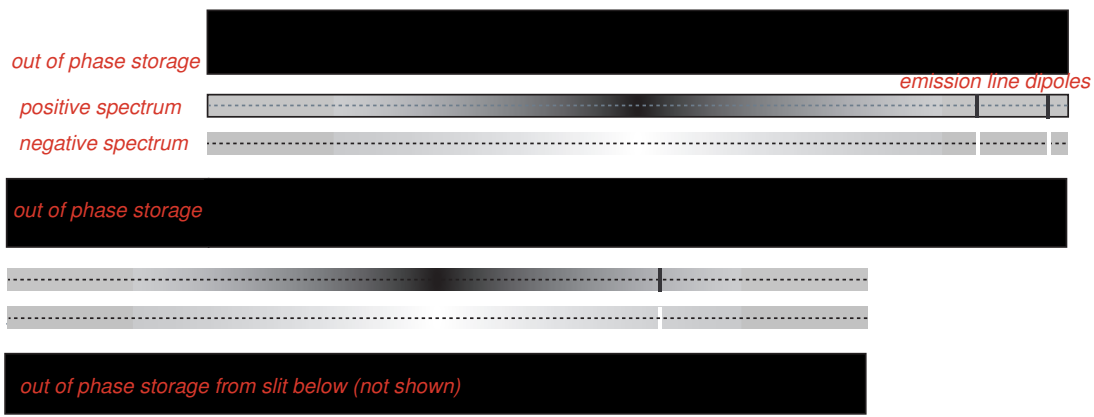} 
\caption{
  Schematic of the Nod and Shuffle mode used in the GDDS.  When the
  telescope is in the object position, CCD area `A' records a
  spectrum. The `sky' position records the nodded spectrum (in this
  case the telescope has been nodded 1.0 arcsec along the slit
  direction). The area `B' is unilluminated by the mask and serves as
  a storage area for the sky position. The image difference A$-$B
  subtracts the sky, and leaves a positive and negative object
  spectrum for subsequent extraction.}
\label{fig:nod-shuffle}
\end{center}
\end{figure*}

The principles of the Nod \& Shuffle technique are described in
Glazebrook \& Bland-Hawthorn (2001). The basic ideas behind the mode
are very similar to those of the {\em Va et Vient} strategy for sky
subtraction \citep{cui94,bla95}\footnote{The main difference between
  {\em Va et Vient} and Nod \& Shuffle is the latter's emphasis on
  using tiny slits for extreme multiplexing.}. Our specific Nod \&
Shuffle configuration is shown schematically in
Figure~\ref{fig:nod-shuffle}, and corresponds to the `Case 1' strategy
shown in Figure~2(a) of \citet{abr03}. Slits were 0.75 arcsec wide
(giving a spectral FWHM of $\simeq 17$\AA) and were designed to take
advantage of the queue observing mode of the telescope by being
optimized for seeing $<0.85$ arcsec. In this seeing the target
galaxies were mostly unresolved. In Nod \& Shuffle mode the telescope
is nodded between two positions along the slit denoted `A' and `B', as
shown in Figure~\ref{fig:nod-shuffle}.  For our first mask
observations on the 22$^{\rm h}$ field we used a 2.0 arcsec long slit,
since the nod distance was 1.0 arcsec the targets appear on the slit
in both A and B positions ($\pm 0.5$ arcsec from the slit center). The
shuffle distance was 28 pixels which produces a shuffled B image
immediately below the A image with a small one pixel gap. For
subsequent fields we increased the slit length to 2.2 arcsec and the
shuffle to 30 pixels as analysis of the first field convinced us that
a slightly longer slit would be beneficial to reduce the impact of the
`red end correction' issue described below.

\begin{figure*}[htbp]
\begin{center}
\includegraphics[height=7.9in]{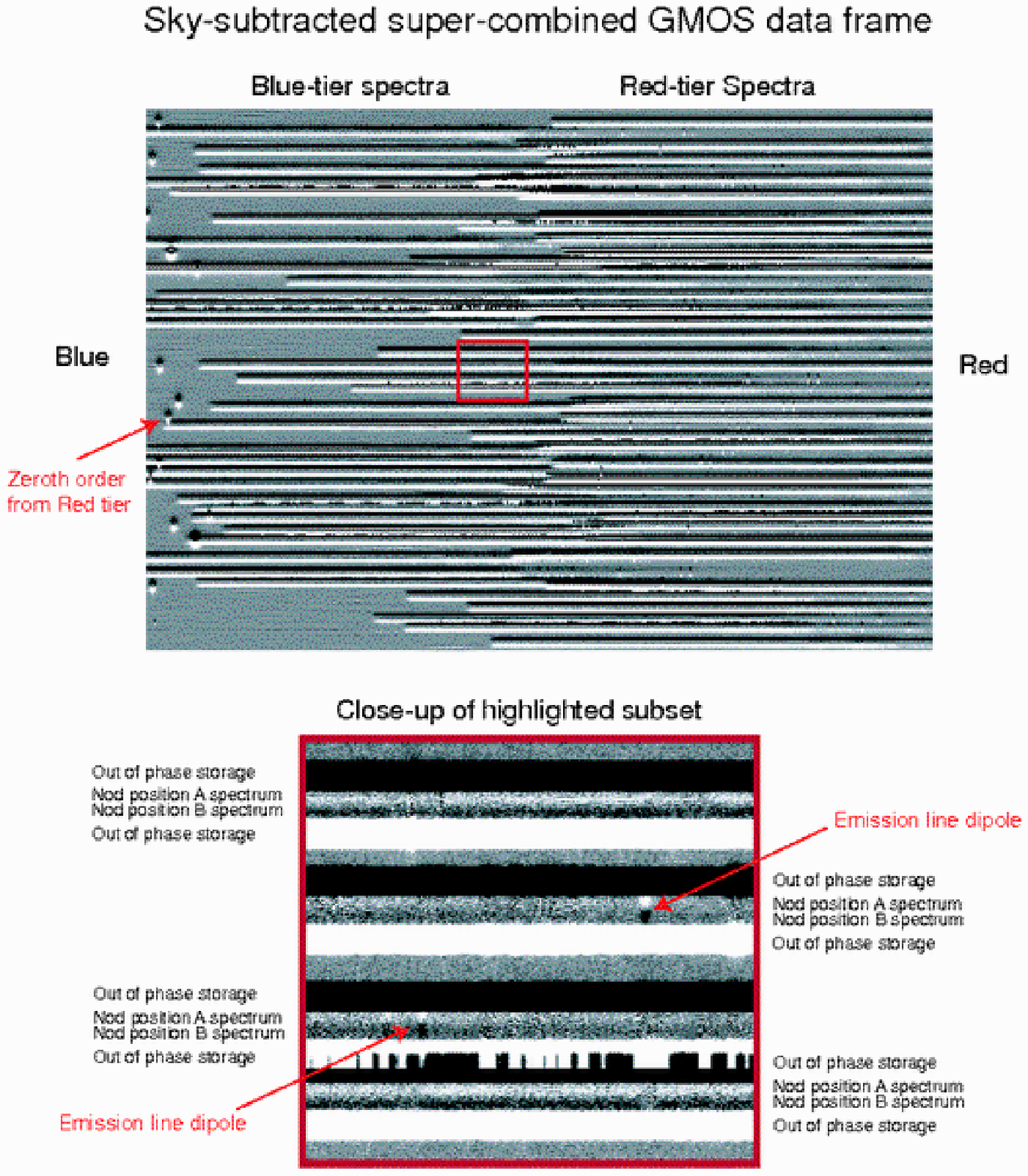} 
\caption{An image showing a sky-subtracted ``supercombined''  frame
  for the GDDS-SA22 field. Object spectra are sandwiched between
  positive and negative sky spectra on the mask. The lower half of the
  figure shows a magnified view of the small rectangular region near
  the center of the mask image. This magnified view shows subsets of
  four spectra with characteristic positive and negative continua and
  emission lines. Note that for clarity we have reversed the GMOS data
  format to show bluer wavelengths to the left.
\label{fig:mask-overview}}
\end{center}
\end{figure*}

At each A and B position we observed for a 60 sec exposure before
closing the shutter and nodding the telescope and shuffling the CCD.
Our standard GDDS exposure consisted of 15 cycles with A$=$60s and
B$=$60s (i.e. a total of 1800s open shutter time on target before
reading out).  There is of course extra overhead associated primarily
with moving the telescope and guide probes, for the GDDS observing
setup this typically added 25\% to the total observing time. We found
that this Nod \& Shuffle setup gave a sky-residual of only
0.05--0.1\%, which is well below the Poisson limit for our stacked
exposures, except for the brightest few night sky lines. The reader is
referred to Abraham et al. (2003) and Murowinski et al 2004 (in
preparation) for a more detailed description of the implementation,
observing sequence and sky-subtraction performance of Nod \& Shuffle
mode on GMOS.

A typical mask observation consisted of approximately 50 half-hour
exposures observed across many nights. In order to fill in the gaps
between CCD chips the grating angle was changed between different
groups of observations in order to dither the MOS image along the
dispersion `X' axis relative to the CCD.  Approximately one third of
the data was taken with a central wavelength of 7380\AA, one third
with 7500\AA\ and the final third with 7620\AA. The CCD gaps were
filled in when the different positions were combined to make a master
frame.

Similarly it is also desirable to dither in the orthogonal, `Y'
direction, i.e. parallel to the spatial axis, in order to minimize the
effects of shallow charge traps aligned with the silicon lattice of the
CCD. These charge traps manifest themselves in the shuffled images as
short pairs of streaks in the horizontal, or dispersion, direction. 
Each pair is always separated by the shuffle distance. We speculate
these charge traps originate from subtle detector defects that are
repeatedly pumped by the shuffle-and-pause action. Since the traps
always appear at the same place on the CCD, their undesired effect can
be greatly reduced by dithering the image along the Y axis and rejecting
outliers during stacking. To accomplish this dithering the CCD was
physically moved using the Detector Translation Assembly (DTA). During
normal GMOS operation, this stage is used to actively compensate for
flexure in the GMOS optical chain during exposures. Additional offsets
can be applied between exposures in order to position the image on
different pixels on the array.

Our standard observing block thus consisted of the following six-step
sequence:
\begin{enumerate}\item The grating was set to one of the 3 positions used (e.g. 7500\AA). The DTA was homed.
\item An 1800s exposure was recorded using Nod \& Shuffle (A$=$60s, B$=$60s, $\times$ 15 cycles).
\item An 1800s exposure was recorded with the DTA offset by +41\micron\  (+3 pixels) along the spatial axis.
\item An 1800s exposure was recorded with the DTA offset by +81\micron\  (+6 pixels) along the spatial axis.
\item An 1800s exposure was recorded with the DTA offset by $-$41\micron\  ($-$3 pixels) along the spatial axis.
\item An 1800s exposure was recorded with the DTA offset by $-$81\micron\  ($-$6 pixels) along the spatial axis.
\end{enumerate}
This same sequence would then be repeated for the next grating
position, and this pattern of changing the central wavelength and
taking a sequence of 5 exposures was repeated until the required
number of exposures was completed. In practice not every sequence had
this exact number of steps and sequence due to observing constraints,
but an approximate balance was maintained among the different DTA
offset positions which was all that was required for the stacking of
the data.

\section{REDUCTION OF TWO-DIMENSIONAL DATA}   
\label{sec:datared}

The goal of the 2D reduction was to combine all the individual 2D
dispersed 1800s exposures for each mask with outlier rejection to make
a master `supercombine' 2D sky-subtracted dispersed image. This is
then used for the next stage --- extraction to 1D spectra. The GDDS 2D
data was reduced using IRAF and the Gemini IRAF package, in particular
the v1.4 GMOS sub-package.  To handle the peculiarities of Nod \&
Shuffle data two new software tasks ({\tt gnsskysub} and {\tt
  gnscombine}) were written by us. These have now been incorporated
into the standard IRAF GMOS package distributed by the Gemini
Observatory.

The first step in the 2D data reduction was to bias subtract the
individual runs using a master bias frame (the average of typically 20
bias frames taken during the observing period). GMOS
exhibits 2D bias structure so a 2D bias subtraction is done with the
{\tt gireduce} task.

The next step was to sky-subtract all the runs using the {\tt gnsskysub}
task. This simply takes the frame, shifts it in Y by the shuffle step,
as recorded in the image header, and subtracts it from
itself\footnote{In some rare cases the detector controller would drop a
sub-frame resulting in an A$\ne$B mismatch. {\tt gnsskysub} includes an
option to fix this case via re-scaling of the B frame}.  This results in
clean sky-subtracted spectra sandwiched between regions of artifacts, as
shown in Figure~\ref{fig:nod-shuffle}. These subtracted frames are then
visually examined to make a list of relative dither offsets. In most
cases these are as given by the nominal DTA offsets, with occasional
1--2 additional pixel shifts between different GMOS nights (these shifts
are then recovered by sky-line fitting and cross-correlation).  In some
cases the objects moved slightly in the slits relative to their nominal
position due to an error in setting the tracking wavelength in the
telescope control system. In order to handle these extra offsets actual
emission lines in bright galaxies were centroided in X and Y and used to
define the offsets. Calculating the offsets directly relative to the
object positions in this way results in some fuzziness of slit edges in
the 2D combined frames, but since the inter-slit offsets were only a few
pixels and only a few frames were affected, this did not turn out to
be a serious problem in practice. The final result of the inspection is
a list of X,Y offsets between the objects in different dispersed images.

Once the offsets are known image combination proceeds with the {\tt
  gnscombine} task which generates sky-subtracted frames using {\tt
  gnsskysub} and combines them using a variance map calculated from
the median count level in the non-subtracted frames and the known
readout noise and Poisson statistics. Outliers (cosmic rays and charge
traps) are rejected using a 7$\sigma$ cut and retained data is
averaged. The outlier rejection was checked visually by comparing
frames combined with and without rejection and it was verified that
only genuine outliers were rejected.  A median 2D sky frame is also
produced which is used for later wavelength calibration and further
noise estimates.

We first combined the frames in 3 groups according to the central
wavelength using {\tt gnscombine}, i.e.  a combined frame was produced
for each of the 7380\AA, 7500\AA\  and 7620\AA\ positions. At this point
the combined frames are a single Multi-Extension FITS (MEF) file where
each extension represents a separate GMOS CCD as a 2D image. The next
step is to use the {\tt gmosaic} task to assemble the three images for
each group into a single contiguous image using the known geometric
relationship between the three CCDs. {\tt gmosaic} was used in the mode
where the assembly was done to the nearest pixel; no re-sampling or
interpolation scheme was used in order to preserve pixel independence in
the noise map.  Since the data is 4$\times$ over-sampled this does not
result in any significant degradation. Inspection of the {\tt gmosaic}'d
images showed the spectral continuity across the CCD gaps was good to
$\pm 1$ pixel and subsequent wavelength calibration showed the
positioning in the dispersion direction was of a similar accuracy.

Finally the three {\tt gmosaic}'d combined frames for the 7380\AA,
7500\AA\  and 7620\AA\ positions were mosaiced again into the final
`supercombine' frame. The task {\tt gemcombine} was used to accomplish
this with offsets calculated from fitting to sky lines. A binary mask
denoting the position of CCD gaps and bad columns was also used to
remove these features from the final supercombine by replacing them with
real data from the other frames.

The final products of this process are: (a) a supercombined (i.e. sky
subtracted) frame corresponding to the the stack of all the 2D data
--- an example is shown in Figure~\ref{fig:mask-overview}; (b) a
corresponding supercombined sky frame showing the emission from the
night sky. It should be noted that no attempt was made to flat-field
the data. This is not required to get accurate sky-subtraction with
Nod \& Shuffle. The effect of pixel-pixel variations in the final
object spectra are greatly reduced by the extensive dithering in any
case.  Some residual flat-field features, primarily fringes, are
visible in the brightest spectra at the few percent level, but these
do not seriously impact our faint spectra.

\section{EXTRACTION OF ONE-DIMENSIONAL SPECTRA}
\label{appendix:oned}

One-dimensional spectra were extracted from the two-dimensional
stacked image frames using {\tt iGDDS}, a publicly available spectral
extraction and analysis program for Mac OS X that we have written for
use with Nod \& Shuffled GMOS data.  {\tt iGDDS} operates in a manner
that is rather different from the familiar command-line driven tools
used by astronomers (e.g. {\tt IRAF, FIGARO, MIDAS}, etc.), and it is
intended to be highly interactive and take full advantage of the
graphical capabilities of modern computers. The program functions as
an electronic catalog with interactive tools for spectral aperture
tracing, one-dimensional spectrum extraction, wavelength calibration,
spectral template fitting, and redshift estimation.  All these tools
are linked. For example, selecting an object in a catalog displays its
two-dimensional image and corresponding aperture trace. This aperture
trace can be reshaped by dragging with a mouse, resulting in a newly
extracted one-dimensional spectrum. Clicking the mouse on a feature on
this spectrum immediately displays this feature on a corresponding
two-dimensional image. This feature can then be selected and a trial
redshift assigned, which results in the superposition of a comparison
template spectrum on the object spectrum. The template can then be
dragged with the mouse to refine the redshift or try other
possibilities.

All analysis steps for all spectra on a GMOS mask are stored in a
single document file which can be shared with colleagues and
interactively modified. The saved {\tt iGDDS} document files used by
our team are publicly available, and the interested reader may find
these to be a useful starting point for further exploration of the
GDDS data set.

\subsection{Aperture Tracing}

Spectra were extracted from the supercombined stack using aperture
traces defined by a fourth-order bezier curve. Positive and negative
apertures (corresponding to the A and B nod and shuffled positions)
were defined relative to this curve. The sizes of these apertures and
the distance between them were allowed to vary independently. (As
described in \S\ref{sec:collisions}, in some cases it is useful to
discard a single A or B position channel in order to avoid
contamination from overlapping spectra). In most cases the spectra
were sufficiently bright to allow a well-defined trace to be
determined visually (after lightly smoothing the two-dimensional image
and displaying the galaxy spectrum with a large contrast stretch).
However, the very faintest galaxies in our sample proved too dim to
allow a reliable trace to be estimated, and for these objects simple
horizontal trace was used.

\begin{figure*}[htb]
\centering \includegraphics[height=4.3in,angle=0]{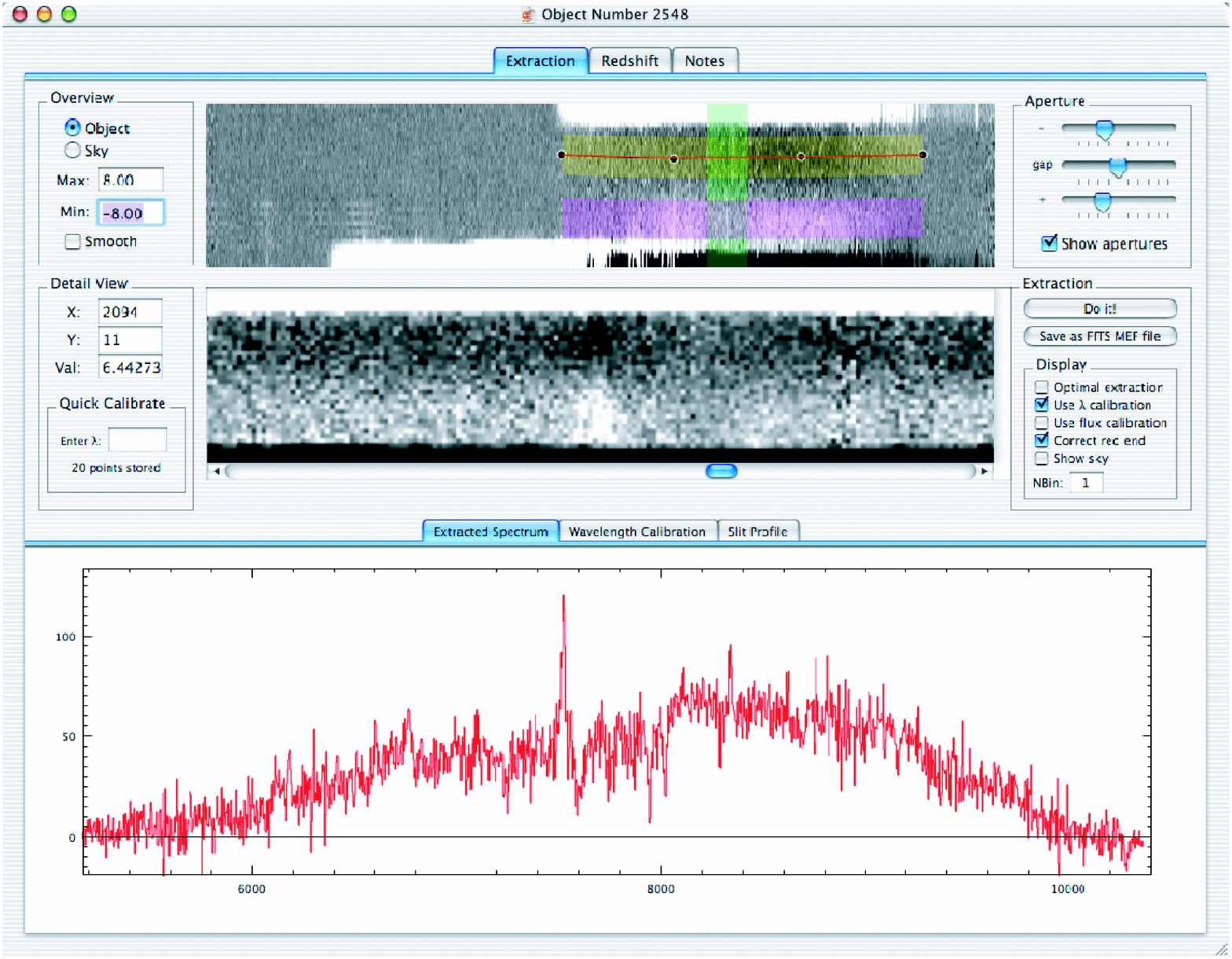}
\caption{An example showing an extraction window in {\tt iGDDS} for
  object 2548 in the $22^{\rm h}$ GDDS field. Two sub-images and the
  corresponding one-dimensional spectrum are shown. The top sub-image
  shows a compressed view of a horizontal slice (34 pixels high and
  4608 pixels wide) across the supercombined 2D image at the
  $y$-coordinate of the slit. The bezier curve defining the trace for
  the spectrum is shown as the solid red line in the top sub-image.
  Control points for the curve are shown as black circles attached to
  the endpoints of the curve. Dragging a control point with the mouse
  changes the shape of the trace. The transparent yellow and purple
  regions linked to the trace correspond to the negative and positive
  extraction apertures. The transparent green rectangle in the top
  sub-image shows the portion of this image that is magnified for
  detailed inspection in the lower sub-image. The extracted spectrum
  for this object is shown at the bottom of the window. Note that at
  this stage in the extraction the spectrum is not fluxed.}
\label{fig:iGDDSExtraction}
\end{figure*}  

A screenshot from {\tt iGDDS} illustrating the aperture tracing
process is shown in Figure~\ref{fig:iGDDSExtraction}. In this figure
the top sub-image shows a compressed view of a horizontal slice across
the supercombined 2D image at the $y$-coordinate of the slit (34
pixels high and 4608 pixels wide). The fourth-order bezier curve
defining the trace for an individual spectrum is shown as the solid
red line in the top window. The transparent yellow and purple regions
on this image are linked to the trace and correspond to the negative
and positive apertures used to extract the one-dimensional spectrum.

\subsection{Linear and Optimal Extraction}

One-dimensional spectra were extracted from the supercombined images
using both linear and optimal extraction procedures. Both sets of
extractions are available in the public data release of the GDDS
observations described in \S\ref{sec:format} below.

Our optimal extractions were constructed using profile weights defined
by projecting the spectra in the spatial direction following the
bezier curves which define their traces. This procedure resulted in
smooth extraction profiles that resembled gaussians for most galaxies,
although in cases where spectrum overlaps occurred (described in
greater detail \S\ref{sec:collisions}) the resulting weight profiles
were spuriously asymmetric. The optimal extractions should not be
trusted for these objects. {\em We therefore recommend that those
  readers interested in making uniform comparisons between all spectra
  in the GDDS use only the linearly extracted spectra}. Optimally
extracted spectra can be trusted for those objects with recorded
spectrum overlap classifications of zero in Table~3, and these do have
slightly improved signal-to-noise relative to their linearly extracted
counterparts. However, the signal-to-noise improvement is modest (of
order 5\%) on account of the narrow slit lengths in the GDDS masks.
For the sake of consistency, only linearly extracted spectra have been
shown in the figures throughout the present paper.

Small artifacts on the two-dimensional supercombined images were
masked out using {\tt iGDDS} prior to extraction. Our masking
procedure works by excluding aberrant pixels from the resulting
average over the spatial direction in a given column. The procedure is
clearly of rather limited usefulness, and was only adopted in those
cases where at most a few pixels in a given column were contaminated
by artifacts. In cases of more severe contamination, we chose to
eliminate the column completely (leaving gaps in the spectrum) rather
than to patch over the bad data. As will be described in
\S\ref{sec:format}, the output data format for the GDDS spectra
retains a record of which wavelength points on a spectrum have been
patched.

Including the effects of bad pixel masking, extraction proceeded as
follows. Consider a single column in a two-dimensional spectrum
containing $n$ rows. Denote the flux in the $i$th pixel by $F_i$, its
variance by $\sigma_i^2$, and
let the discrete variable $\mathcal{M}_i \in \{-1,0,1\}$ take on the
value $1$ in the case that the pixel is in aperture A, $-1$ in the
case that the pixel is in aperture B, and $0$ in the case that the
pixel is masked. Assuming $n_A$ pixels are contained within apertures
A and B, a simple estimator of the total flux in the case of linear
extraction with masked regions is:

\begin{equation}
F = n_A \cdot  {\sum_{i=1}^n (\mathcal{M}_i/\sigma_i^2) F_i \over \sum_{i=1}^n | (\mathcal{M}_i/\sigma_i^2) |} 
\end{equation}

\noindent In other words, the flux is now the average over the 
non-rejected pixels multiplied by the total number of masked and
un-masked pixels in both apertures.

The corresponding case for optimal extraction is only slightly more
complicated. Denoting the optimal extraction profile by a continuous
variable $P_i \in \{0..1\}$ (with $P_i \equiv 0$ for pixels in the
column outside the aperture), it is straightforward to show that the
maximum likelihood estimator for the true total flux is given by:

\begin{equation}
F = \sum_{i=1}^n P_i \cdot { \sum_{i=1}^n P_i  (\mathcal{M}_i/\sigma_i^2) F_i \over \sum_{i=1}^n   P_i^2 | (\mathcal{M}_i/\sigma_i^2) | }
\end{equation}

\noindent Note that if we set $P_i = 1$ then we recover the linear extraction case. 

\subsection{Flux Calibration, Atmospheric Absorption Correction, and Red Fix Correction}
\label{sec:calib}

Since precise flux calibration is impossible for observations
accumulated over many nights (spread over months in some cases) under
varying conditions, the flux calibration was carried out using
observations of standard stars obtained as part of the GMOS queue
baseline calibration. These data were reduced using the standard
routines in IRAF and calibrations deduced for each field. A mean
aperture correction of a factor of 3.5 was applied to each spectrum.
In the end, the calibrations were so similar from field to field that
the same one was used for all. For most objects the relative flux
calibration appears to be quite good, as evidenced by the fact that
composites made from these spectra agree extremely well with
composites from other surveys, e.g., the SDSS luminous red composite
of \citet{eis03}.  However, the absolute flux for an individual object
may be in considerable error (we think they should only be trusted to
within about a factor of two), since no attempt was made to allow for
overlapping spectra, masked regions, flat-fielding of 2D spectra, etc.
We therefore caution that the fluxes are relative only.

As the 1D spectra were initially being extracted from the 2D spectra,
it almost immediately became apparent that there was a problem in that
the continua became too low or even negative at the extreme red end of
the wavelength region. Since one expects that the nod and shuffle
technique would result in perfect sky subtraction, this was initially
very puzzling. On further examination it became evident that the
strong sky lines displayed ``tails" so that there was spatial
extension of the lines that increased in strength with wavelength.
Unfortunately, in the nod and shuffle technique, the protrusion of
these on either side of the spectra means that they are superimposed
and subtracted from the object spectrum during the shift-and-combine
operation. The origin of this effect is charge diffusion in the
silicon which is a strong function of wavelength and it has only
become apparent since we are attempting to extract extremely faint
target spectra that are stored immediately adjacent to extremely
strong sky spectra on the CCD. The effect could be reduced or avoided
by increasing the distance between the object and sky or the two
object positions in our case, at the cost of less efficient use of the
detector area. For any given mask design, the magnitude of the effect
depends on the precise relative position of the object in the slitlet
but, in our case, the only practical way to correct for it was to
establish an average correction for all objects in a mask and then
apply that to all spectra.

A correction curve to account for this effect was derived empirically
for each mask. First, the variation of the strength of the effect as a
function of wavelength was derived by measuring the percentage of light
that leaked from strong sky lines into an extraction window equivalent
to that used for the objects but on the opposite, unexposed, side of
spectrum. It was immediately obvious that the effect varies
exponentially with wavelength, ranging (for our initial mask) from 0.2\%
at 883nm to 5\% at 1030nm. Having established the form of the variation,
it was multiplied by a high signal-to-noise sky spectrum that was
broadened slightly in wavelength to account for the fact that the charge
diffusion occurs in all directions (technically, the broadening should
also be a function of wavelength but this was ignored). Finally, this
modified sky spectrum was scaled so that it minimized the negative sky
features that were apparent in a co-added (in observed wavelength)
spectrum of the 25 strongest spectra in the mask. This `redfix'
correction curve can then be applied as an option in {\tt iGDDS} during
the data reduction. As mentioned above, it is at best only a statistical
correction for objects in a mask, and this effect introduces additional
uncertainty into the continuum level and flux calibration that decreases
exponentially shortward of 1 micron. The magnitude of the effect also
increases with the faintness of the target since it is a fixed fraction
of the night sky.

Our final spectra have also been corrected for the major features
caused by atmospheric absorption.  The atmospheric features were
identified by isolating them in a normalized, high signal-to-noise
spectrum of a standard star and then adjusting their amplitude to
match those in the combined spectrum of suitable objects in a given
mask.

\subsection{Spectrum Order Overlaps}
\label{sec:collisions}

\begin{figure*}[htb]
\centering
\includegraphics[height=5in,angle=0]{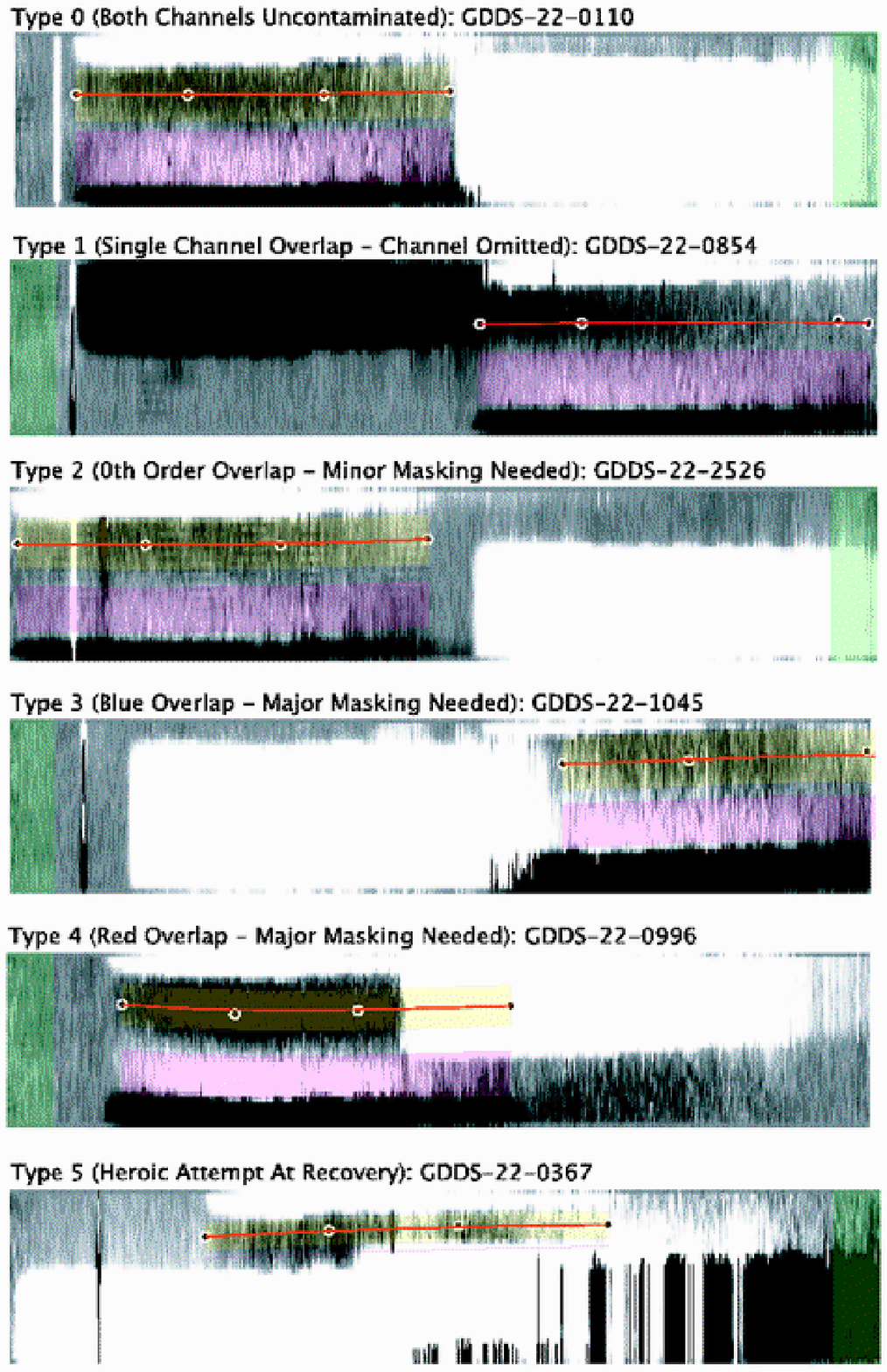}
\caption{Examples illustrating the spectrum overlap classification
  system defined in Table~3. The transparent yellow and purple regions
  correspond to the negative and positive spectral extraction
  apertures.}
\label{fig:overlap}
\end{figure*}

As described in \S\ref{sec:design}, our two-tier mask design strategy
allowed some overlap to occur among spectra originating in adjacent
slits. The extent of these overlaps can be gauged by an inspection of
Figure~\ref{fig:mask-overview}. The wavelengths 5500--9200\AA\ from
the `blue tier' spectra in second order can overlap with the first
order `red tier' spectra. Since 5500--9200\AA\ is the main
observational window, this second order light cannot be filtered out,
nor is the second order sky cancelled in the Nod \& Shuffle process,
because the slits are in general not aligned between the two tiers.
However the intensity of second order spectra in this wavelength range
is only 5-10\% of the intensity of the first order spectra, so in
practice only the very strongest sky lines (such as [OI]5577\AA,
[OI]6300\AA~ and [OI]6363\AA) were significant contaminants, and in
many cases these individual lines can simply be masked out, as
described above.  Another source of contamination is zero-order light
from red-tier spectra overlapping with the first-order spectra in the
blue tier. Several examples of this contamination are clearly seen in
Figure~\ref{fig:mask-overview}. Such cases are easy to identify and,
as only a small portion of the spectrum is affected, this portion of
the affected spectra has simply been eliminated in the final spectra
presented in this paper. As noted earlier, we have attempted to
classify the importance of spectrum overlaps using the system defined
in Table~\ref{tab:collision}. To help the reader visualize the meaning
of this system, it is illustrated using example spectra in
Figure~\ref{fig:overlap}.

\subsection{Final ASCII-format Spectra}
\label{sec:format}

In addition to the main data tables presented in Tables 4 and 5 above,
the GDDS Public Data Release contains the individual spectra for all
galaxies in the survey. These spectra are stored as ASCII text files,
each of which contains the eleven columns of information specified in
Table \ref{tab:dataformat}. Linearly and optimally extracted
calibrated spectra, their corresponding variance spectra, and an
uncalibrated raw spectrum are stored in the same file. A linearly
extracted night sky spectrum sampled through the same slit as part of
the Nod \& Shuffle operation is also included. All calibrated spectra
have been fully processed through our pipeline; they are
flux-calibrated, as well as corrected for atmospheric absorption and
charge bleeding (via the `redfix' correction described in
\S\ref{sec:calib}). Separate columns in the output data file record
the values of the atmospheric calibration, `redfix', and flux
calibration curves applied at each wavelength point in the spectrum
(so the calibrations can be undone and new ones experimented with). A
column in each file also records the fraction of masked pixels in the
spatial direction at each wavelength. The final column in each file
records raw counts in electrons normalized to the 1800s exposure time
of a single sub-frame, as described above.

\begin{deluxetable}{lll} 
\tablenum{6}
\tablecolumns{3} 
\tablewidth{0pc} 
\tabletypesize{\small} 
\tablecaption{Data Columns in Final Text-Format Spectrum Files\label{tab:dataformat}}
\tablehead{ 
  \colhead{Name} & 
  \colhead{Quantity} &
  \colhead{Unit}
} 
\startdata
{\tt Lambda} & Wavelength& \AA \\
{\tt Flux\tablenotemark{a}} & Linearly extracted object flux & erg ${\rm cm}^{-2} {\rm s}^{-1} {\rm \AA}^{-1}$\\
{\tt  Sigma\tablenotemark{a}} & Standard deviation of linearly extracted object flux & erg ${\rm cm}^{-2} {\rm s}^{-1} {\rm \AA}^{-1}$\\
{\tt  SkyFlux\tablenotemark{a}} & Linearly extracted sky flux & erg ${\rm cm}^{-2} {\rm s}^{-1} {\rm \AA}^{-1}$\\
{\tt  OptFlux\tablenotemark{a}}  & Optimally extracted object flux & erg ${\rm cm}^{-2} {\rm s}^{-1} {\rm \AA}^{-1}$\\
{\tt  OptSigma\tablenotemark{a}} & Standard deviation of optimally extracted object flux & erg ${\rm cm}^{-2} {\rm s}^{-1} {\rm \AA}^{-1}$\\
{\tt  RedFix} & Additive correction for charge bleeding & counts \\
{\tt  FluxCa}l & Flux calibration & mag \\
{\tt Atmos} & Additive correction for atmospheric absorption & counts  \\
{\tt  Frac} & Fraction of pixels masked in spectral dimension & \nodata \\
{\tt  Electrons} & Uncorrected counts & electrons \\
\enddata
\tablenotetext{a}{Flux-calibrated and corrected for atmospheric absorption and charge bleeding. As
described in the text (see \S C.3), the absolute flux calibration for individual objects is only approximate.}
\end{deluxetable}

\end{document}